\documentclass[conference]{IEEEtran}

\PassOptionsToPackage{bookmarks=false}{hyperref}

\usepackage{cite}
\usepackage{hyperref}
\hypersetup{hidelinks}

\usepackage{orcidlink}
\usepackage{amsmath,amssymb,amsfonts}
\usepackage{algorithmic}
\usepackage{algorithm}
\usepackage{array}
\usepackage[caption=false,font=footnotesize]{subfig}
\usepackage{textcomp}
\usepackage{stfloats}
\usepackage{url}
\usepackage{verbatim}
\usepackage{graphicx}
\usepackage{balance}
\usepackage{xcolor}
\usepackage{tabularx}
\usepackage{booktabs}
\usepackage{balance}

\def\BibTeX{{\rm B\kern-.05em{\sc i\kern-.025em b}\kern-.08em
    T\kern-.1667em\lower.7ex\hbox{E}\kern-.125emX}}
\begin{document}

\title
{Modeling Maritime Transportation Behavior Using AIS Trajectories and Markovian Processes\\in the Gulf of St. Lawrence}

\author{
\centering
\IEEEauthorblockN{
Gabriel Spadon\IEEEauthorrefmark{1},
Ruixin Song\IEEEauthorrefmark{2},
Vaishnav Vaidheeswaran\IEEEauthorrefmark{1}}
\IEEEauthorblockN{
Md Mahbub Alam\IEEEauthorrefmark{1},
Floris Goerlandt\IEEEauthorrefmark{2},
Ronald Pelot\IEEEauthorrefmark{2}
}

\IEEEauthorblockA{\IEEEauthorrefmark{1}\textit{Faculty of Computer Science}, \textit{Dalhousie University}, \textit{Halifax}, \textit{Canada}}
\IEEEauthorblockA{\IEEEauthorrefmark{3}\textit{Dept. of Industrial Engineering}, \textit{Dalhousie University}, \textit{Halifax}, \textit{Canada} \\
\{spadon, rsong, vaishnav, mahbub.alam, floris.goerlandt, ronald.pelot\}@dal.ca}
}

\maketitle

\begin{abstract}
Maritime transportation is central to the global economy, and analyzing its large-scale behavioral data is critical for operational planning, environmental stewardship, and governance. This work presents a spatio-temporal analytical framework based on discrete-time Markov chains to model vessel movement patterns in the Gulf of St. Lawrence, with particular emphasis on disruptions induced by the COVID-19 pandemic. We discretize the maritime domain into hexagonal cells and construct mobility signatures for distinct vessel types using cell transition frequencies and dwell times. These features are used to build origin-destination matrices and spatial transition probability models that characterize maritime dynamics across multiple temporal resolutions. Focusing on commercial, fishing, and passenger vessels, we analyze the temporal evolution of mobility behaviors during the pandemic, highlighting significant yet transient disruptions to recurring transport patterns. The methodology we contribute to this paper allows for an extensive behavioral analytics key for transportation planning. Accordingly, our findings reveal vessel-specific mobility signatures that persist across spatially disjoint regions, suggesting behaviors invariant to time. In contrast, we observe temporal deviations among passenger and fishing vessels during the pandemic, reflecting the influence of social isolation measures and operational constraints on non-essential maritime transport in this region.
\end{abstract}


\section{Introduction}

Maritime mobility is fundamental to international trade, directly influencing the global economy~\cite{unctad2023rmt} and environmental stewardship. Vessel movements, such as commercial shipping, fishing operations, and passenger activities, significantly impact transportation safety, ecological conservation, and infrastructure planning~\cite{paolo2024satellite}. Given the growing complexity of marine traffic systems, systematically analyzing mobility patterns is key for improving transport operations, optimizing routing, and supporting sustainable governance of our oceans~\cite{sharif2024multi}.

The Gulf of St. Lawrence in eastern Canada is a region where diverse vessel types meet, including commercial ships, fishing vessels, and passenger boats. Due to heavy maritime traffic, this region is exposed to operational risks and ecological pressures~\cite{harvey1999preliminary}. Understanding vessel behavior patterns in this area has direct implications for transport strategy, economic resilience, and environmental risk mitigation.

Automatic Identification System (AIS) datasets provide real-time tracking data on vessel positions, speeds, and courses, making them essential for large-scale monitoring and transport analytics~\cite{Stach2023Maritime}. These datasets enable detailed spatiotemporal analysis of vessel movements. However, the volume and variability of AIS data still demand advanced data-driven methods to extract structured knowledge and support modeling~\cite{fu2020ais, Dalaklis2023}.

AIS has been integral to Canada's maritime transport safety and environmental protection since its adoption in international conventions~\cite{imo2000ais}. The Navigation Safety Regulations mandated that most domestic vessels operate an AIS transceiver to support navigational monitoring and situational awareness~\cite{canada2019navigation}. The Canadian Coast Guard has built a network of shore-based AIS stations to observe traffic in national waters~\cite{fournier2018past}. AIS data have enabled studies focused on reducing the impact and increasing spatial coordination of marine activities~\cite{spadon2024probabilistic, hermannsen2019recreational}.

Accordingly, this paper presents an AIS-based analytical framework that applies stochastic modeling to characterize vessel mobility as a transport system in the Gulf of St. Lawrence. To support structured analysis, we quantize maritime space into hexagonal cells and convert continuous vessel trajectories into discrete transition sequences. These transitions are used to compute mobility signatures for each vessel type, which are then modeled using spatial transition probability matrices derived from discrete-time Markov chains.

We utilize this framework to assess the operational impact of the COVID-19 pandemic on maritime transportation activities. During this period, travel restrictions and operational constraints affected transport systems worldwide, providing a unique opportunity to observe behavioral adaptation and systemic vulnerability. Our results show that certain vessel types maintained consistent mobility signatures over time, indicating stable operational patterns across space. Others, particularly fishing and passenger vessels, exhibited marked disruptions and spatial redistribution, reflecting differences in functional roles and exposure to service-level policies.

This study contributes a methodological approach for analyzing large-scale maritime transport behavior using probabilistic modeling and temporal segmentation. It provides insight into how different vessel categories respond to exogenous disruptions and reveals stable patterns that can inform transport resilience assessment and decision support. Accordingly, the main contributions of this work are summarized as follows:
\setlength{\itemsep}{0pt}
\setlength{\parskip}{0pt}
\begin{itemize}
\item We propose an analytical framework that models vessel mobility patterns as discrete-time Markov processes for spatio-temporal characterization of vessel behavior.
\item We present a set of mobility metrics derived from Markovian models to capture both spatial and temporal aspects of maritime dynamics across vessel types.
\item We perform a large-scale, multi-year empirical study of vessel mobility to uncover vessel-specific behavioral invariants and their resilience to external disruptions.
\item We identify and quantify pandemic-induced deviations in vessel mobility, demonstrating the differential sensitivity of vessel categories to socio-economic shocks.
\end{itemize}

The remainder of this paper is structured as follows: Section~\ref{sec:related_works} reviews prior research on spatio-temporal mobility modeling, maritime traffic analysis using AIS data, and pandemic-induced behavioral changes. Section~\ref{sec:methodology} details our proposed analytical framework, including the spatial discretization process, Markovian modeling approach, and temporal segmentation strategy. Subsequently, Section~\ref{sec:results} presents a comprehensive analysis of vessel mobility patterns, highlighting vessel-type-specific behavioral invariants and pandemic-related disruptions. Finally, Section~\ref{sec:conclusions} summarizes our findings and discusses implications for transport planning and monitoring.

\section{Related Works}
\label{sec:related_works}

Understanding mobility patterns through probabilistic modeling has been a central research focus across domains, including human movement prediction and maritime transportation analytics~\cite{lu2013approaching, song2010modelling, liu2020universal}. In particular, Markovian frameworks have demonstrated strong performance in modeling spatio-temporal dependencies and uncovering latent behavioral structures in complex dynamic systems~\cite{Saputra2024Mobility, farnoosh2020deepmarkov}. Recent efforts have extended these models by incorporating auxiliary information and scaling their application to large trajectory datasets, such as those derived from AIS records~\cite{alam2025physics, wenzhe2025stgdpm, nguyen2024, yu2025multimodal}. These advances support increasingly granular modeling of transport behavior across space and time. Additionally, the disruptions caused by the COVID-19 pandemic have allowed researchers to study the resilience of transport systems under exogenous shocks, revealing differential impacts across transport modes and operational contexts. The following studies describe important contributions that motivated our analytical approach.

Yan~{\it et al.} (2021)~\cite{Yan2021Mobility} propose a weighted Markov chain for predicting user mobility based on cellular network data. By classifying users according to the complexity of their trajectories, the authors train specialized models per group. Their contribution lies in demonstrating improved predictive accuracy over baseline Markov chains, emphasizing the relevance of personalized modeling for heterogeneous behavior.

Shi~{\it et al.} (2024)\cite{Shi2024Combining} combine Tucker decomposition with Mobility Markov Chains to model spatiotemporal mobility. Their technique captures latent dependencies across heterogeneous modes and regions, outperforming classical Markovian methods and offering potential applications in predictive transport analytics.
Similarly, Xia{\it et al.} (2023)~\cite{Xia2023Community} apply discrete-time Markov chains with Dirichlet regression to analyze daily human activity trajectories. Integrating demographic and environmental features enables their model to reveal interpretable activity patterns at the community level, supporting population-scale transport planning and behavioral profiling.

Kim~{\it et al.} (2022)\cite{Kim2022Maritime} employ spatio-temporal density analysis over AIS datasets to assess maritime traffic distribution. Their work identifies key shipping corridors and quantifies dynamic congestion patterns, contributing to improved navigation safety and maritime infrastructure optimization. Likewise, March{\it et al.} (2021)~\cite{March2021Tracking} investigate the early-stage impact of COVID-19 on global maritime mobility using AIS data. They observe significant reductions in vessel activity, especially among passenger ships, and provide a quantitative basis for understanding transport system responses to emergencies.

Loveridge~{\it et al.} (2024)~\cite{Loveridge2024Context} evaluate pandemic-induced variability in vessel operations by sector. Using AIS records spanning multiple months, they identify enduring reductions in passenger traffic and regional increases in fishing activity, emphasizing asymmetric adaptation within the maritime domain. On different approach, Wang~{\it et al.} (2022)~\cite{Wang2022Quantitative} analyze AIS-derived port activity metrics to assess COVID-19 impacts on logistics infrastructure. Their findings include elevated anchoring and berthing durations, as well as increased near-port vessel concentrations, underscoring the operational strain placed on transport nodes during global disruptions.

To the best of our knowledge, no prior work has jointly applied a discrete-time Markov modeling framework with mobility- and dwell-time-based transport metrics to characterize vessel behavior in the Gulf of St. Lawrence using AIS data. This study contributes by analyzing vessel mobility over a multi-year period and systematically quantifying the resilience of the transport system during the COVID-19 pandemic. Specifically, we identify vessel-type-specific behavioral invariants and measure disruption sensitivity across commercial, fishing, and passenger categories. These findings reveal spatial redistribution patterns and operational signatures, offering actionable insights for traffic management, transport infrastructure planning, and resilience assessment in a geographically complex and environmentally sensitive region.

\section{Methodology}
\label{sec:methodology}

We use satellite-based AIS data~\footnote{Spire Maritime -- \url{https://spire.com/maritime/}} spanning the period from 2013 to 2023 to model vessel mobility in a transport systems context. The dataset contains key navigational attributes, including timestamp, latitude, longitude, speed over ground, course over ground, and navigational status. We preprocess the raw AIS records by resampling vessel trajectories into standardized time intervals. Afterward, we discretize the maritime space into a spatial grid and map these resampled trajectories to corresponding grid cells. Using this representation, we extract aggregated movement patterns and derive behavioral metrics through probabilistic modeling, which captures latent dynamics and structural patterns in regional maritime traffic.

\begin{table}[!b]
    \caption{Table of Notations.}
    \label{tab:notation}
    \begin{tabularx}{\linewidth}{@{}>{\centering\arraybackslash}p{1.6cm}X@{}}
        \toprule
        \textbf{Symbol} & \textbf{Description} \\
        \midrule
        $\tau_v$ & Raw AIS trajectory of vessel $v$ (ordered). \\
        $T$ & Ordered set of observation timestamps. \\
        $(x_t,y_t)$ & Latitude and longitude recorded at time $t$. \\
        $\Delta t$ & Uniform resampling interval ($1\,\mathrm{min}$). \\
        $\mathcal{S},\,s_i$ & Hexagonal state grid; $s_i$ is one cell. \\
        $\widetilde{\tau}_v$ & Resampled sequence of $(s_i,t)$ pairs for $v$. \\
        $X_t$ & Random state (cell) occupied at step $t$. \\
        $p_{ij},\,P$ & One-step transition probability and its matrix. \\
        $\mathcal{N}(i)$ & Set of neighbors directly reachable from $s_i$. \\
        $\mathcal{D}_{ij}$ & Dwell durations in $s_i$ before $s_i\!\!\to\!s_j$. \\
        $N_{ij}$ & Count of $s_i\!\!\to\!s_j$ transitions. \\
        $w_{ij}$ & Mean dwell in $s_i$ before exiting to $s_j$. \\
        $\lambda_{ij}$ & Hazard rate $1/w_{ij}$. \\
        $q_{ij},\,Q$ & Dwell-weighted transition probability and its matrix. \\
        $\pi_i$ & Stationary probability of occupying $s_i$. \\
        $\mathrm{MM}_{i}$ & Mobility magnitude $\sum_j N_{ij}$. \\
        $\mathrm{DTM}_{i}$ & Dwell-time magnitude $\sum_j N_{ij}w_{ij}$. \\
        $\sigma_{jk},\sigma_{jk}(i)$ & Number of shortest paths $s_j\!\to\!s_k$; subset via $s_i$. \\
        $C_i$ & Betweenness centrality of state $s_i$. \\
        $d_{ij}$ & Length of the shortest path $s_i\!\to\!s_j$. \\
        $k_i$ & Strength of state $s_i$: $\sum_j R_{ij}$. \\
        $c_i$ & Community label of state $s_i$. \\
        $\delta$ & Kronecker delta. \\
        $\mathcal{D}_{w}$ & Raw distribution of node weights (for clipping). \\
        $Q_{\text{low}},Q_{\text{high}}$ & Lower/upper percentile thresholds. \\
        $\mathfrak{S}$ & Spline transform applied after clipping. \\
        $n$ & Number of states: $|\mathcal{S}|$. \\
        \bottomrule
    \end{tabularx}
\end{table}

\subsection{Trajectory Representation}

We define a trajectory as a sequence of spatial positions recorded over discrete time intervals. Given a vessel $v$, its trajectory over an observation window is represented as:
\begin{equation}
    \tau_v \;=\; \bigl\{ (x_t,\,y_t) \bigr\}_{t \in T},
\end{equation}
where $(x_t, y_t)$ is the latitude and longitude recorded at timestamp $t$, and $T = \{t_1 < t_2 < \dots < t_{|T|}\}$ is the set of observation timestamps. Due to irregular AIS transmission intervals, raw trajectories exhibit variable temporal resolution~\cite{Zhang2016}. To standardize trajectories for comparative analysis, we segment each trajectory into 3-hour windows and resample positional data at fixed time intervals $\Delta t$ (set to $1~\text{min}$) using linear interpolation. The standardized trajectory is then given by:
\begin{equation}
    \widetilde{\tau}_v \;=\; \bigl\{ (x_{t'},\,y_{t'}) \bigr\}_{t' \in T'},
\end{equation}
where $T' = \{t'_1, t'_2, \dots, t'_{|T'|}\}$ denotes the set of uniformly spaced timestamps. This temporal normalization enables the construction of a discrete-state Markovian model for characterizing mobility across vessel types, periods, and regions.

\subsection{State-Space Definition}

We transform continuous trajectories into discrete sequences by partitioning the maritime domain into a finite set of states using a uniform hexagonal grid. In comparison to conventional square tiling, hexagonal binning provides uniform neighbor connectivity and reduces directional bias~\cite{sahr2011hexagonal}, improving spatial fidelity in mobility modeling. To define the spatial state space, we use the hierarchical hexagonal indexing system \textbf{\textit{H3}}\footnote{\url{https://h3geo.org/}}, which divides the Earth's surface into equally spaced hexagonal cells~\cite{Wozniak2021hex2vec}. For mobility and dwell-time computations, we select resolution level 6, which produces approximately $8{,}687$ hexagonal cells, each covering around $36\,\mathrm{km}^{2}$.

Each coordinate $(x_t, y_t)$ from the resampled trajectory is mapped to a corresponding spatial cell $s_i \in \mathcal{S}$, where $\mathcal{S} = \{s_1, s_2, \ldots, s_n\}$ denotes the set of states. The trajectory $\widetilde{\tau}_v$ is thus expressed as a discrete-time sequence of visited states:
\begin{equation}
    \widetilde{\tau}_v = \{(s_{t_1}, t_1), (s_{t_2}, t_2), \dots, (s_{t_{|T|}}, t_{|T|})\}.
\end{equation}

\subsection{Markovian Modeling}
\subsubsection{The first-order assumption}

We model vessel mobility as a stochastic process in which the probability of transitioning to the next state depends only on the current state (i.e., memoryless dynamics). Prior studies in human and maritime mobility have shown that first-order Markov chains achieve a favorable balance between model complexity and representational fidelity~\cite{Guo2018kChain, Kulkarni2019}, motivating our approach. Formally, the vessel’s position at time $t+1$ satisfies the Markov property:
\begin{align}
    \mathbb{P}(X_{t+1} = s_j \mid\ &X_t = s_i, X_{t-1}, \dots, X_0) \nonumber \\
    &= \mathbb{P}(X_{t+1} = s_j \mid X_t = s_i),
\end{align}
where $X_t \in \mathcal{S}$ is the state occupied at time $t$. This enables construction of a one-step transition matrix $P = [p_{ij}]$ with:
\begin{equation}
    p_{ij} = \mathbb{P}(X_{t+1} = s_j \mid X_t = s_i), \quad \text{where} \quad \sum_{j \in \mathcal{S}} p_{ij} = 1.
\end{equation}

\subsubsection{Dwell as a behavioral driver}

To further characterize transport behavior, we incorporate dwell-time modeling, which quantifies the duration that vessels remain in a state before transitioning. This is particularly relevant for capturing activities such as anchoring, fishing, or loitering~\cite{Asanjarani2021}. Under the first-order assumption, let $\mathcal{D}_{ij}$ be the set of observed dwell durations in state $s_i$ before transitioning to $s_j$. Let $N_{ij} = |\mathcal{D}_{ij}|$ be the number of transitions, and the mean dwell-time be as:
\begin{equation}
    w_{ij} = \frac{1}{N_{ij}}\sum_{d_k \in \mathcal{D}_{ij}} d_k.
\end{equation}
Bearing such assumptions, we compute the empirical hazard (transition) rate as the inverse of this mean~\cite{Asanjarani2021, smith2021big}:
\begin{equation}
    \lambda_{ij} = \frac{N_{ij}}{\sum_{d_k \in \mathcal{D}_{ij}} d_k}.
\end{equation}
To incorporate dwell-time into our transition model, we normalize the hazard rates across all outgoing transitions from $s_i$, resulting in a dwell-weighted transition matrix $Q$:
\begin{equation}
    q_{ij} = \frac{\lambda_{ij}}{\sum_{j' \in \mathcal{N}(i)} \lambda_{ij'}}, \quad \text{where} \quad \sum_j q_{ij} = 1,
\end{equation}
with $\mathcal{N}(i)$ denoting the set of reachable neighbors from $s_i$. This matrix $Q$ combines spatial transitions with temporal behavior, enabling the analysis of location-dependent patterns.

\subsection{Markovian Metrics}

To derive insights into vessel mobility using stochastic processes, as part of our methodology, we define a collection of network-based metrics that characterize and interpret navigation patterns. These metrics capture relevant features of maritime mobility, such as hotspots, bottlenecks, recurrent pathways, and topological properties embedded in vessel trajectories~\cite{Liu2021, hu2014detecting}. They consider the geographical system as a spatial/planar graph and analyze the relationships among states to describe the underlying transportation system.

Let $P=[p_{ij}]$ denote the one-step transition matrix constructed over the set of hexagonal cells $\mathcal{S}=\{s_1,\dots,s_n\}$.  
We denote by $\boldsymbol{\pi}=(\pi_1,\dots,\pi_n)$ the stationary distribution satisfying $\boldsymbol{\pi}^{\!\top}=\boldsymbol{\pi}^{\!\top}P$; by $N_{ij}$ the observed number of $s_i\!\!\to s_j$ transitions; by $w_{ij}$ the mean dwell time in $s_i$ preceding transitions to $s_j$; and by $P^{k}$ the $k$-step transition matrix~\cite{Kuntz2021}.

\vspace{.1cm}
\noindent\textit{\textbf{Local System Metrics}}

\subsubsection{Mobility Magnitude (MM)}
This metric aggregates the total outgoing transition count from a state, identifying origin hubs and high-frequency departure zones. Larger values indicate active centers such as ports or shipping lanes, while lower values suggest marginal or pass-through regions.
\begin{equation}
  \mathrm{MM}_{i}= \sum_{j\in\mathcal{S}} N_{ij}.
\end{equation}

\subsubsection{Dwell-Time Magnitude (DTM)}
This metric quantifies the cumulative dwell durations in a state prior to outbound transitions. High values reveal anchoring areas, fishing zones, or delay-prone segments, whereas lower values characterize transit corridors with minimal stops.
\begin{equation}
  \mathrm{DTM}_{i}= \sum_{j\in\mathcal{S}} N_{ij}\,w_{ij}.
\end{equation}

\subsubsection{Betweenness Centrality (C)}
This measure captures the extent to which a state acts as a connector on shortest paths between all other state pairs. High centrality indicates structural choke points or interchange hubs; low values correspond to peripheral or isolated nodes~\cite{Doglioli2017BetweennessMarine}.
\begin{equation}
  C_i = \sum_{j \neq i} \sum_{k \neq i,\,j}
        \frac{\sigma_{jk}(i)}{\sigma_{jk}},
\end{equation}
where $\sigma_{jk}$ is the number of shortest paths from $s_j$ to $s_k$, and $\sigma_{jk}(i)$ counts those passing through $s_i$.

\vspace{.1cm}
\noindent\textit{\textbf{Global System Metrics}}

\subsubsection{Average Path Length ($\mathcal{L}$)}
This metric calculates the average number of transitions along the shortest paths between all reachable state pairs. It serves as a global proxy for network compactness. Lower values reflect a well-connected maritime domain, while higher values suggest sparse patterns~\cite{Cullinane2016PortCentrality}.
\begin{equation}
  \mathcal{L}=\frac{1}{n(n-1)}\sum_{i\ne j} d_{ij},
\end{equation}
where $n=|\mathcal{S}|$ and $d_{ij}$ is the shortest path length from $s_i$ to $s_j$ in the directed graph induced by $p_{ij}>0$.

\subsubsection{Modularity ($\mathcal{Q}$)}
This metric evaluates the extent to which the mobility network clusters into internally dense yet externally sparse communities. It highlights sub-regions where vessels tend to circulate within a local area. A high modularity suggests localized behavioral regimes, while low values reflect more uniform or entangled flows.
\begin{equation}
  \mathcal{Q}=\frac{1}{2y}\sum_{i,j}
      \left(R_{ij}-\frac{k_i k_j}{2y}\right)\,
      \delta\left(c_i,c_j\right),
\end{equation}
where $R_{ij}=\tfrac12(p_{ij}+p_{ji})$ is the edge weight, $k_i=\sum_j R_{ij}$ is the node strength, $y=\tfrac12\sum_{i,j}R_{ij}$ is the total weight, $c_i$ is the community label of $s_i$, and $\delta$ is the Kronecker delta.

\subsection{Pre-Analytics Processing and Strategies}

\subsubsection{Dataset Segmentation}
The full AIS dataset spans 2013 to 2023. For focused spatiotemporal analysis related to the COVID-19 pandemic, we isolate the core period from January 2020 to December 2022. We partition this interval into:
\begin{equation}
\begin{aligned}
    \mathcal{T}_{\text{pre}}           &= \bigl[\text{2019‑01‑01},\,\text{2019‑12‑31}\bigr],\\
    \mathcal{T}_{\text{pandemic}_{P1}} &= \bigl[\text{2020‑01‑01},\,\text{2020‑12‑31}\bigr],\\
    \mathcal{T}_{\text{pandemic}_{P2}} &= \bigl[\text{2021‑01‑01},\,\text{2021‑12‑31}\bigr],\\
    \mathcal{T}_{\text{post}}          &= \bigl[\text{2022‑01‑01},\,\text{2022‑12‑31}\bigr].
\end{aligned}
\end{equation}
The pre-pandemic window $\mathcal{T}_{\text{pre}}$ offers a behavioral reference. The $\mathcal{T}_{\text{pandemic}_{P1}}$ window covers the strictest public-health measures in Canada, initiated on March 15, 2020~\cite{PHAC2020Isolation}. The $\mathcal{T}_{\text{pandemic}_{P2}}$ window captures partial relaxation, while $\mathcal{T}_{\text{post}}$ reflects post-pandemic normalization. For vessel-specific analysis, we disaggregate data into commercial $\mathcal{V}^{(c)}$, fishing $\mathcal{V}^{(f)}$, passenger $\mathcal{V}^{(p)}$, and aggregated $\mathcal{V}^{(a)}$ categories.

\subsubsection{Globalizing Spatial State Metrics}
To compare mobility across temporal windows, local state-based metrics are aggregated into scalar summaries. For each metric $\boldsymbol{\phi}^{(\ell)}$ in window $\ell$, we compute a stationary-distribution-weighted spatial average:
\begin{equation}
    \Phi^{(\ell)}_{\!\phi}
    = \sum_{i=1}^{n}
      \pi^{(\ell)}_i\,
      \phi^{(\ell)}_i,
    \qquad
    \sum_{i=1}^{n}\pi^{(\ell)}_i = 1,
\end{equation}
where $\pi^{(\ell)}_i$ is the stationary probability of occupying state $s_i$ during window $\mathcal{T}_\ell$.

\subsubsection{Metrics Quantization}
Many stochastic metrics follow heavy-tailed distributions. To reduce the influence of outliers, we first clip raw node-weight distributions $\mathcal{D}_{w}$ to the $[1^{\text{st}}, 98^{\text{th}}]$ percentile range. We then apply a spline transformation $\mathfrak{S}$ to enhance visual contrast in low and mid-value ranges across vessel types. In this case, we consider $Q_{\text{low}}$ and $Q_{\text{high}}$ as the percentile thresholds per vessel type over all windows.
\begin{equation}
    \text{proj}_{[0,1]}(z) = \max(0, \min(1, z))
\end{equation}
Then, for each weight $w$, we compute the transformed value:
\begin{equation}
    \mathcal{\hat{D}} = \mathfrak{S}\left(\text{proj}_{[0,1]}\left(\frac{w-Q_{low}}{Q_{high}-Q_{low}}\right)\right)
\label{eq:clip_and_transform}
\end{equation}

\section{Results}
\label{sec:results}

\vspace{.1cm}
\noindent\textit{\textbf{System‑Level Temporal Dynamics}}
\vspace{.1cm}

\begin{figure}[!h]
    \centering

    \subfloat[State Space -- Node Count -- $|\mathcal{S}|$]{
        \includegraphics[width=.91\columnwidth]{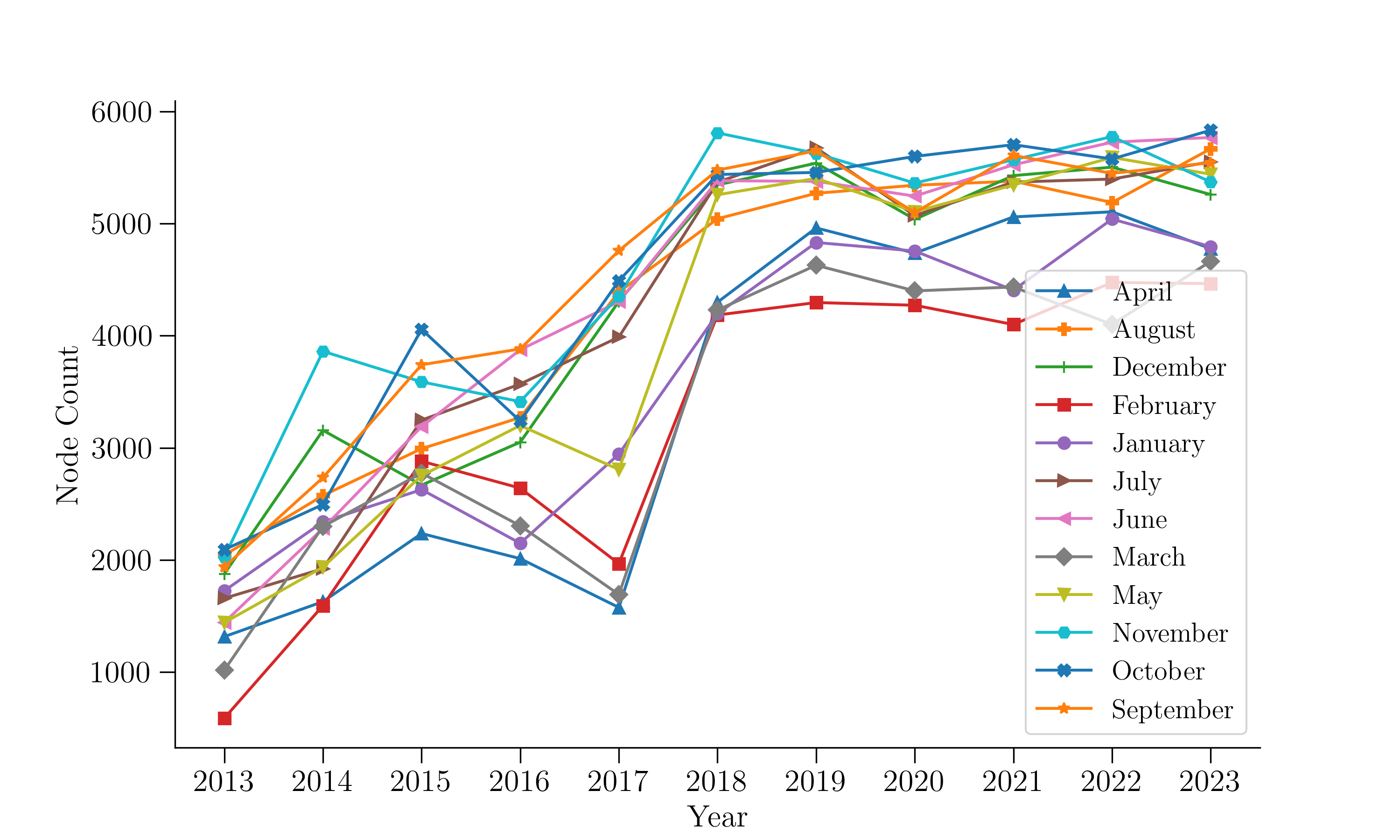}
        \label{fig:node_counts}
    }

    \subfloat[State Space -- Edge Count -- $|\widetilde{\tau}_v|$]{
        \includegraphics[width=.91\columnwidth]{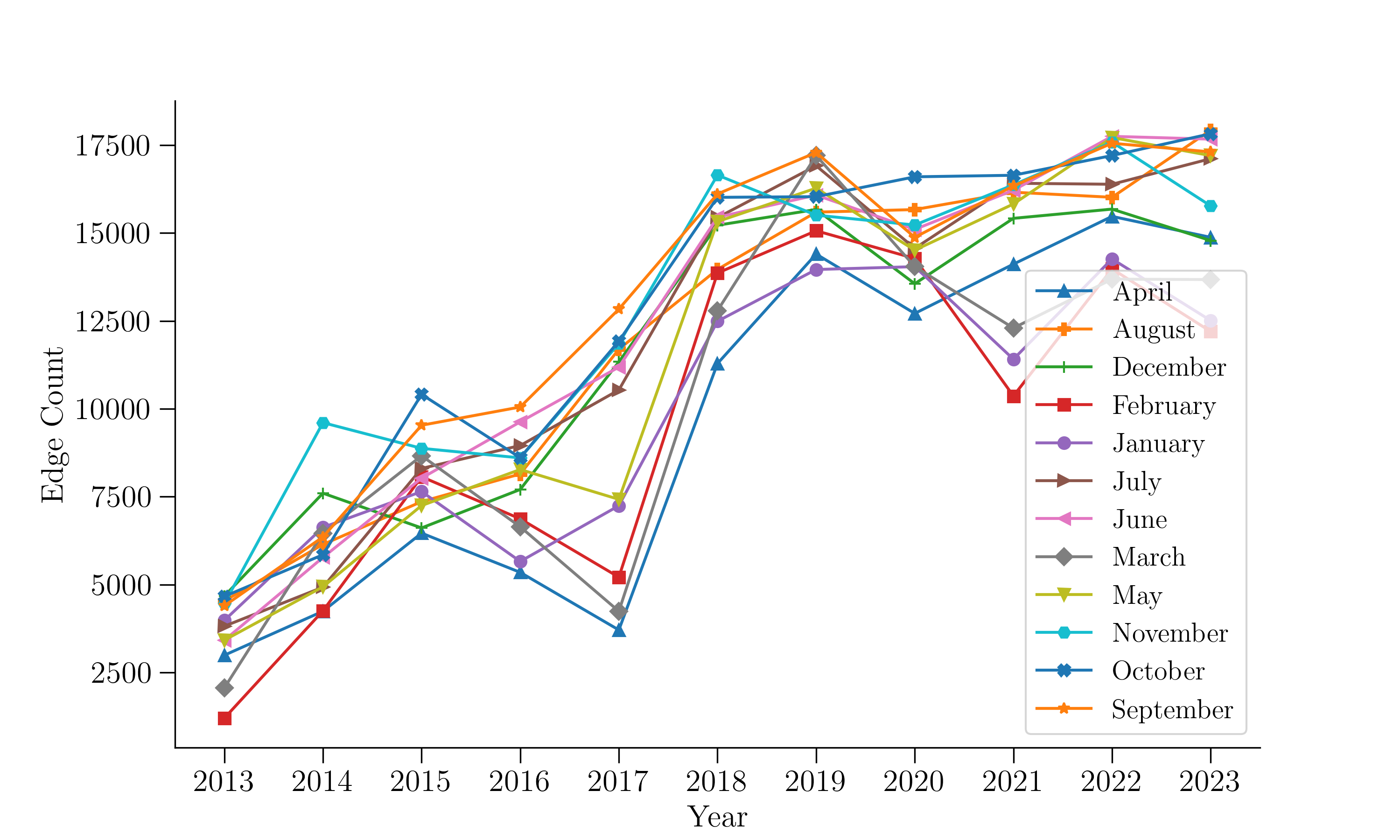}
        \label{fig:edge_counts}
    }

    \subfloat[Modularity -- $\mathcal{Q}$]{
        \includegraphics[width=.91\columnwidth]{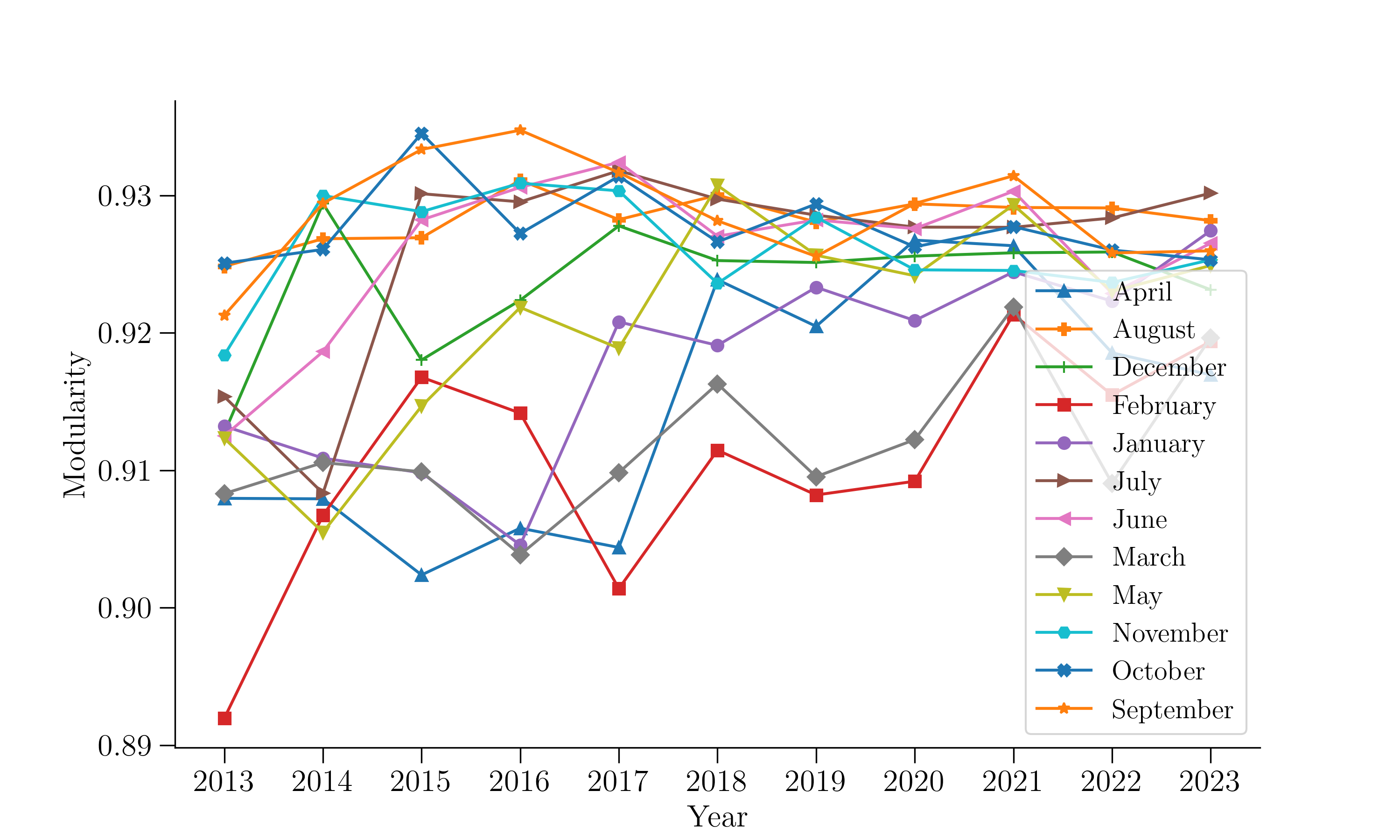}
        \label{fig:modularity}
    }

    \subfloat[Average Path Length -- $\mathcal{L}$]{
        \includegraphics[width=.91\columnwidth]{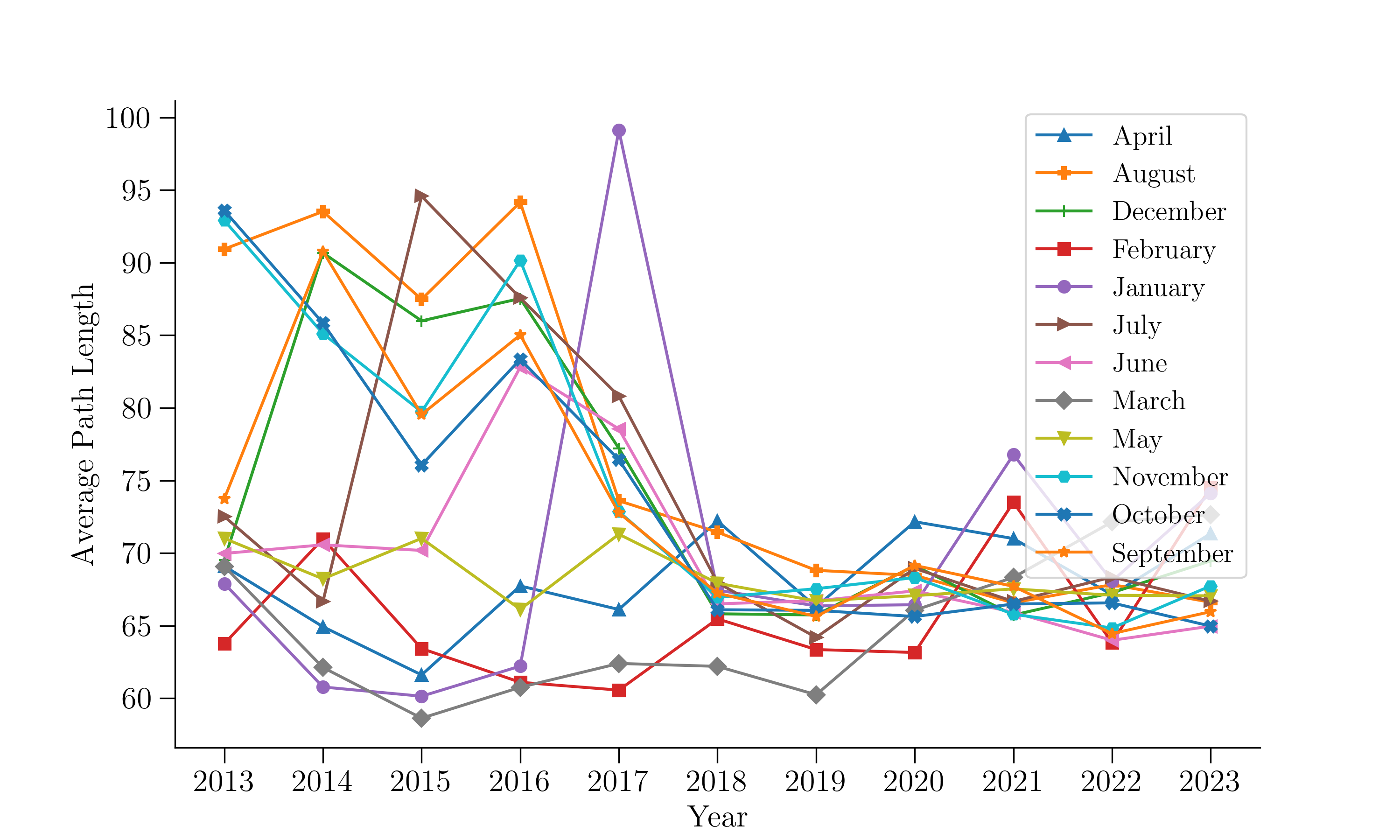}
        \label{fig:avg_path}
    }

    \caption{Global metrics across experiments.}
    \label{fig:all_graph_metrics}
\end{figure}

\begin{figure}[!b]
    \centering

    \subfloat[Betweenness Centrality -- $C_i$]{
        \includegraphics[width=.91\linewidth]{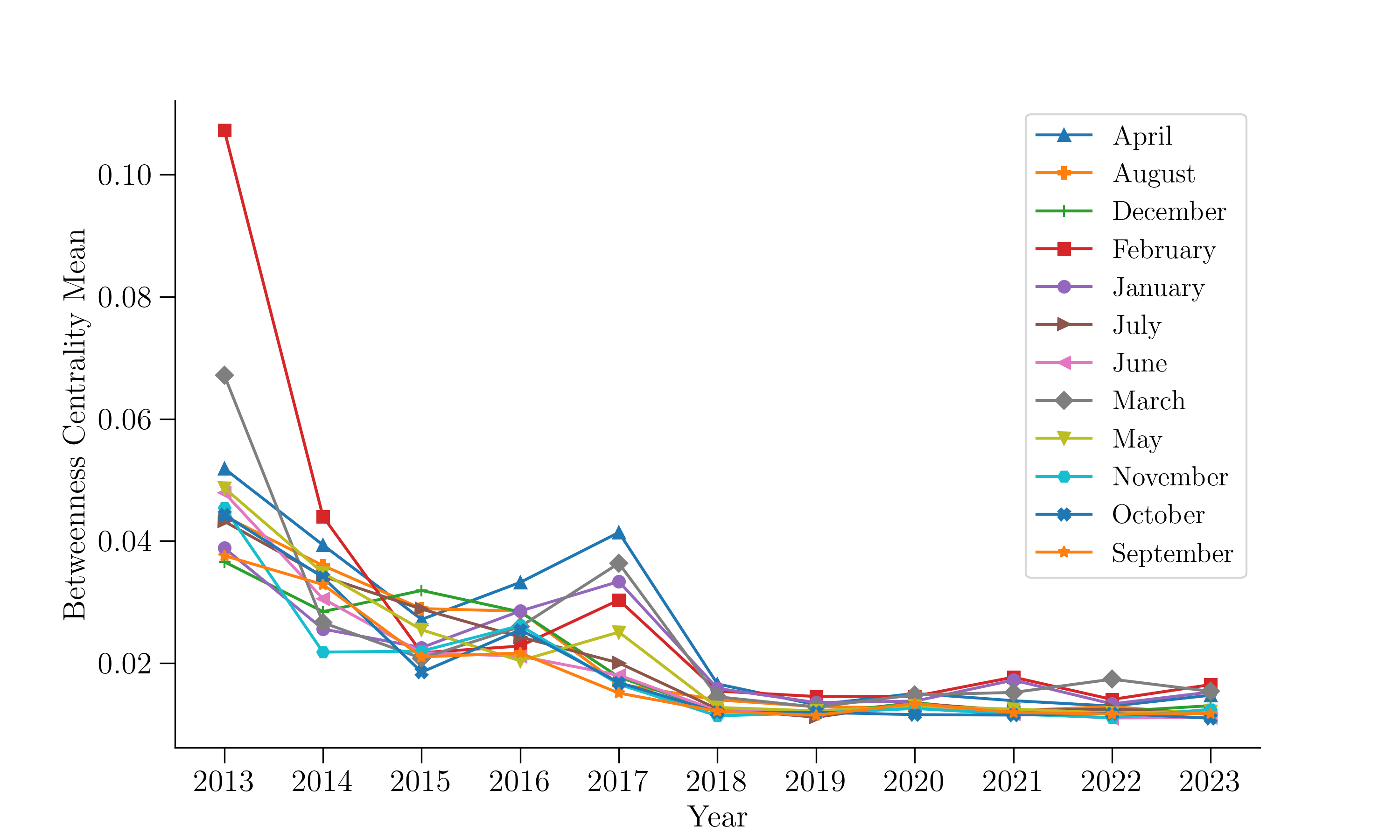}
        \label{fig:betweenness}
    }

    \vspace{0.5em} 

    \subfloat[Mobility Magnitude -- $|\mathrm{MM}_i|$]{
        \includegraphics[width=.91\linewidth]{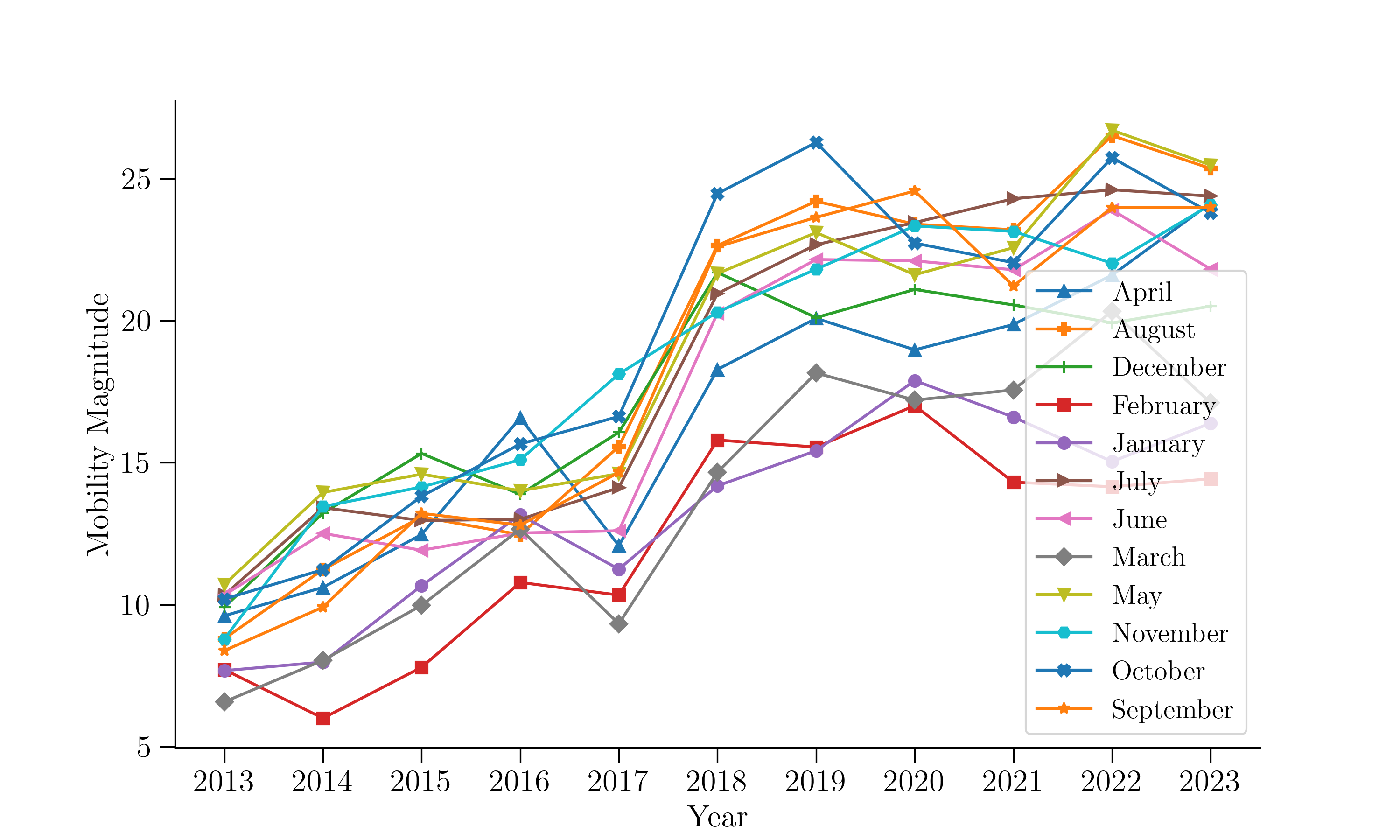}
        \label{fig:mobility_magnitude}
    }

    \vspace{0.5em}

    \subfloat[Dwell Time Magnitude -- $|\mathrm{DTM}_i|$]{
        \includegraphics[width=.91\linewidth]{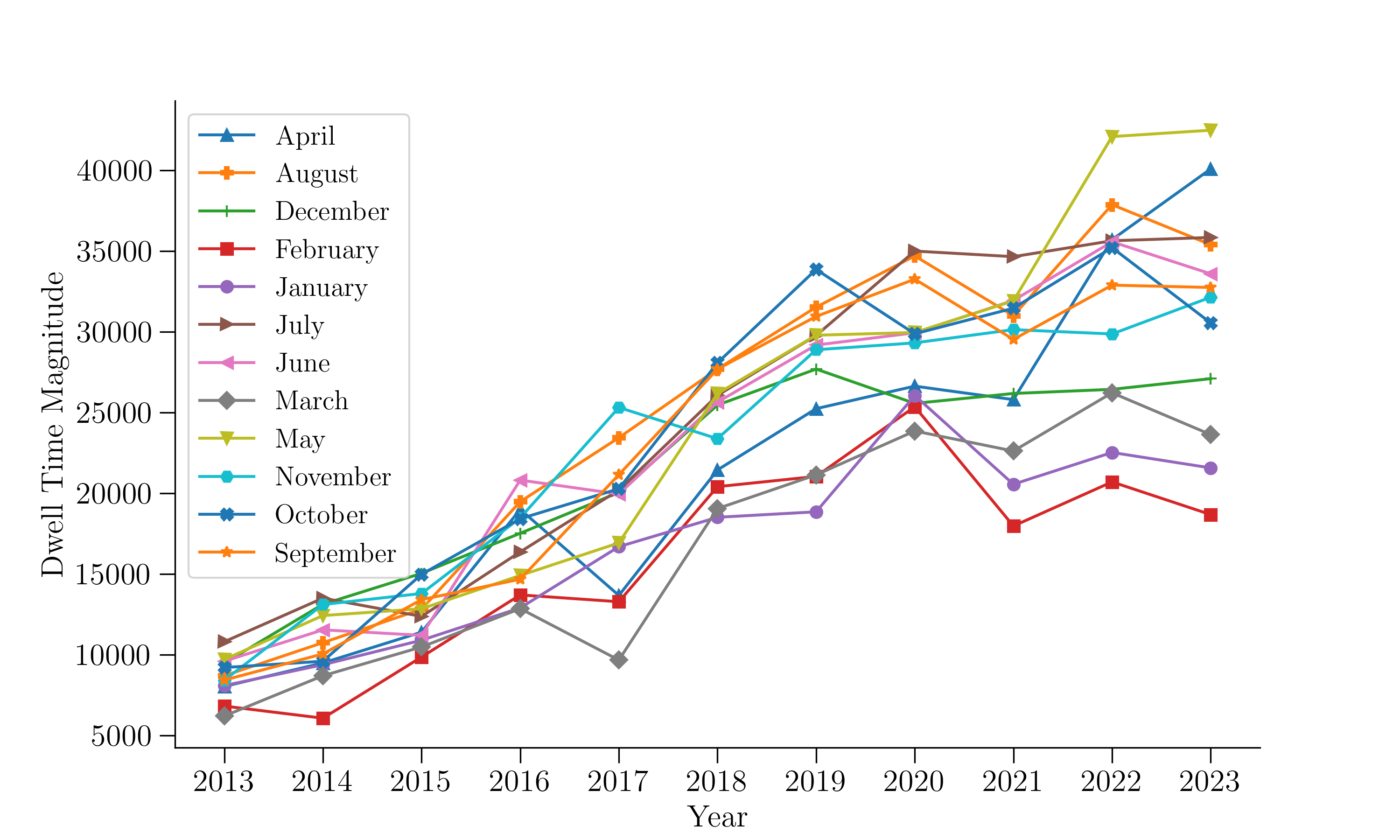}
        \label{fig:dwell_magnitude}
    }

    \caption{Network centrality and system magnitude across experiments.}
    \label{fig:betweenness_and_magnitudes}
\end{figure}

We begin the analysis by examining the evolution of maritime traffic within the study region. Figures~\ref{fig:node_counts} and~\ref{fig:edge_counts} depict the temporal development of the maritime state space from 2013 to 2023. The number of occupied cells $\lvert\mathcal{S}\rvert$ (Figure~\ref{fig:node_counts}) and the number of vessel transitions $\lvert\tilde{\tau}_v\rvert$ (Figure~\ref{fig:edge_counts}) steadily increase until stabilizing around 2018. This trend corresponds to the progressive adoption of AIS technology in Canadian waters. Early records primarily capture sparse traffic from a limited set of vessels navigating the isolated areas of the Gulf of St. Lawrence. As coverage expanded, the spatial extent and connectivity of trajectories grew accordingly.

A significant decline in maritime activity occurred from February to May 2017, followed by a gradual recovery. This temporary downturn aligns with record-setting water levels in the St.~Lawrence River and Seaway during the spring of 2017~\cite{waterflood2017}, which led to speed restrictions and operational constraints on commercial and recreational vessels.

The stabilization of spatial patterns after 2018 suggests that AIS data reached a representative and mature stage, capturing stable navigational dynamics. This aligns with Canada's regulatory requirement for AIS transmitters on domestic vessels implemented in 2019~\cite{canada2019navigation}. Observed fluctuations in state connectivity between 2020 and 2021 likely reflect the effects of the COVID-19 pandemic on ocean mobility behavior.

Figures~\ref{fig:modularity} and~\ref{fig:avg_path} present the annual progression of modularity and average path length over the same spatial and temporal scope. Modularity increases from approximately 0.92 in 2013 to above 0.95 after 2020, suggesting a more sharply defined segmentation of the Gulf into localized traffic basins. Once AIS reporting became comprehensive, internal circulation patterns of shipping, fishing, and passenger vessels appear to have stabilized into consistent structures.

Conversely, the average path length decreases significantly from over 100 transitions in 2013 to around 75 by 2018, followed by stabilization within a narrow range of 72–82. This decline does not indicate any change in the physical geography of the Gulf, but rather reflects an improvement in data resolution through enhanced AIS coverage. As more vessels consistently reported their locations, the observed network grew denser, revealing intermediate transitions that were previously unrecorded. Slight fluctuations after 2019 correspond with pandemic-related operational changes that temporarily altered routing behaviors without modifying the overall navigational topology. The rise in modularity and drop in average path length signal a data system that has become increasingly complete, clustered, and efficiently connected.

Figure~\ref{fig:betweenness} provides further structural context by displaying the monthly evolution of betweenness centrality. The steep decline from 2013 to 2015, followed by a prolonged plateau, reflects a diffusion of network flow across parallel routes as AIS coverage improved. This shift reduced reliance on a few strategic transit nodes. Following 2018, centrality stabilizes at approximately one-quarter of its original level, confirming that no single state monopolizes the shortest-path traffic. The temporary peaks in 2017 and 2021 are seasonal in nature and do not indicate a reemergence of structural chokepoints.

To further assess the dynamic characteristics of maritime activity, Figure~\ref{fig:mobility_magnitude} illustrates the evolution of average \textit{Mobility Magnitude}, which captures the intensity of transitions across cells; Figure~\ref{fig:dwell_magnitude} reports trends in \textit{Dwell Time Magnitude}, reflecting the extent of stationary behaviors within the network.

From Figure~\ref{fig:mobility_magnitude}, we observe a consistent increase in vessel transitions up to 2018, followed by a plateau marked by seasonal variation. August and September consistently exhibit peak traffic volumes, coinciding with the summer season, while mobility levels drop in January and February. The sharp decline in early 2020 aligns with the onset of pandemic restrictions, signaling a clear but temporary disruption in activity. Recovery is expected to progress through 2021, with near-complete restoration by 2022 (typical operational patterns).

Figure~\ref{fig:dwell_magnitude} reveals a more pronounced response. During the pandemic, both mobility and dwell times decline, indicating widespread inactivity and increased anchorage with limited AIS transmission. Even into 2022, dwell remains elevated, suggesting a lag in operational normalization. The seasonal structure persists, with dwell peaks consistently occurring during summer, which is likely attributable to heightened maritime demand during those months.

\begin{figure*}[ht]
    \centering\hspace{-.75cm}
    \setlength{\tabcolsep}{1pt}
    \renewcommand{\arraystretch}{0.9}
    \begin{tabular}{r@{\hspace{4pt}}c@{\hspace{2pt}}c@{\hspace{2pt}}c@{\hspace{2pt}}c@{}}
         \multicolumn{1}{c}{\textit{\textbf{2019}}} &
          \multicolumn{1}{c}{\textit{\textbf{2020}}} &
          \multicolumn{1}{c}{\textit{\textbf{2021}}} &
          \multicolumn{1}{c}{\textit{\textbf{2022}}} \\[1.5pt]
        \includegraphics[width=0.25\linewidth]{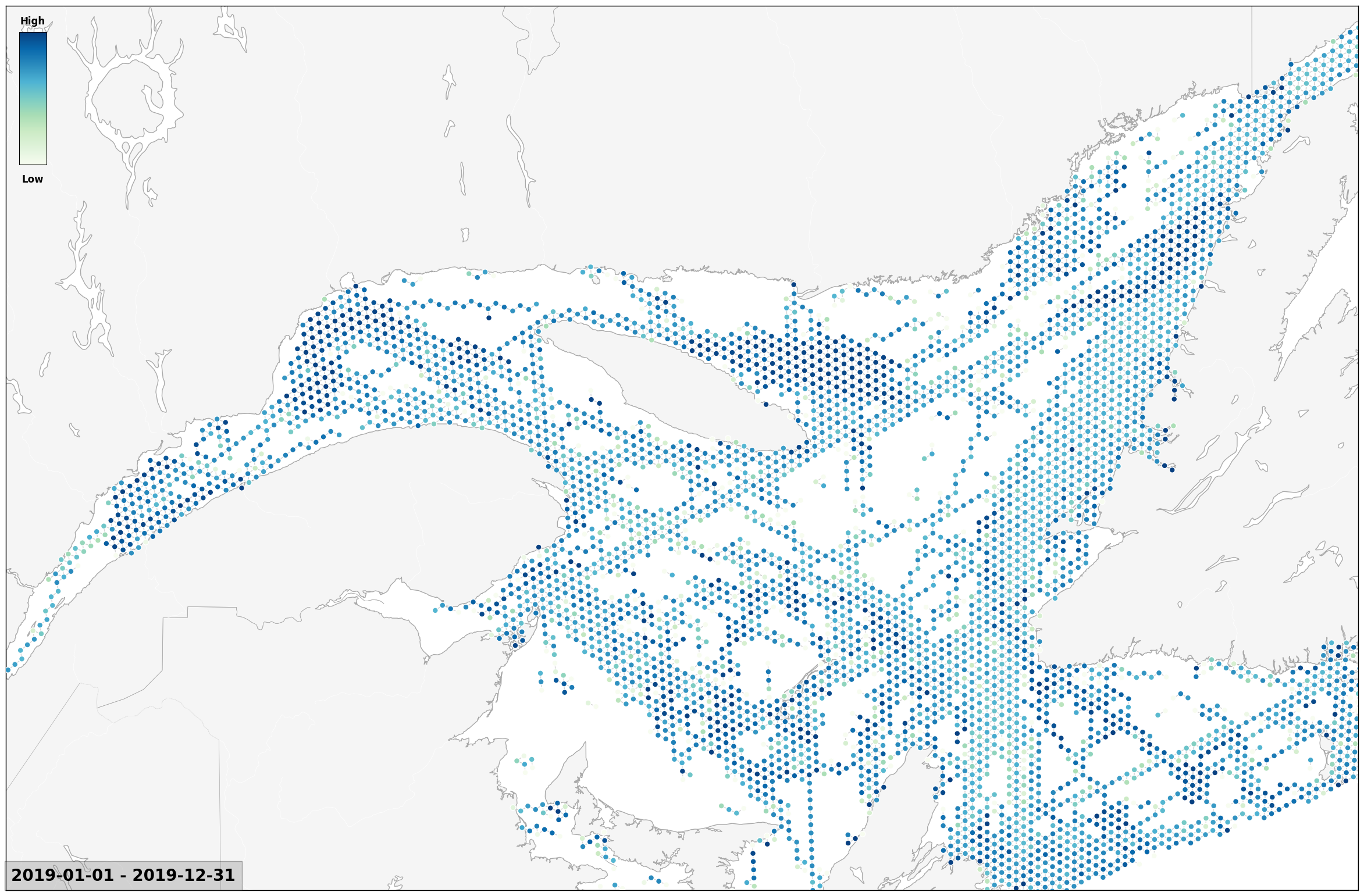} &
        \includegraphics[width=0.25\linewidth]{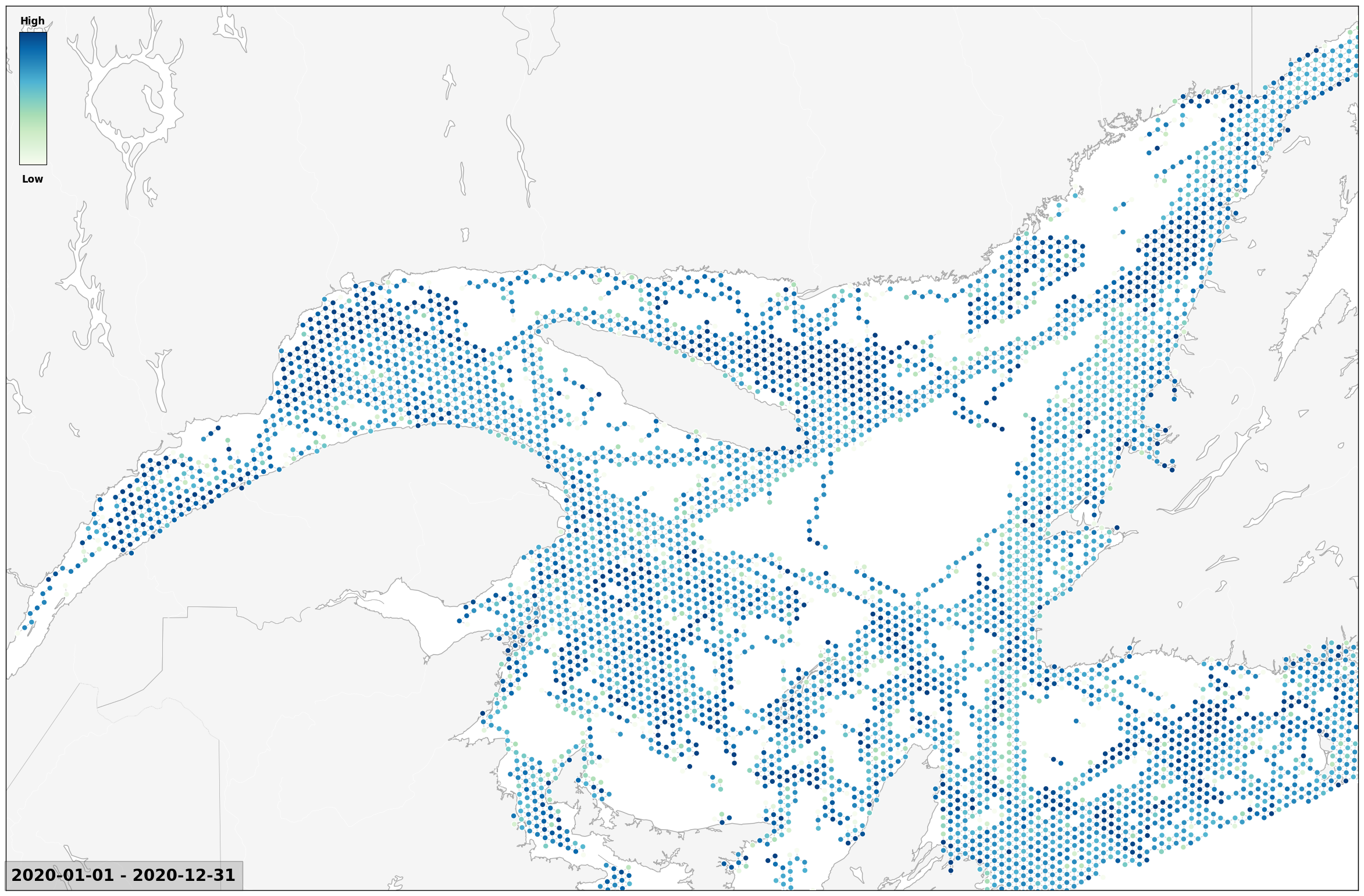} &
        \includegraphics[width=0.25\linewidth]{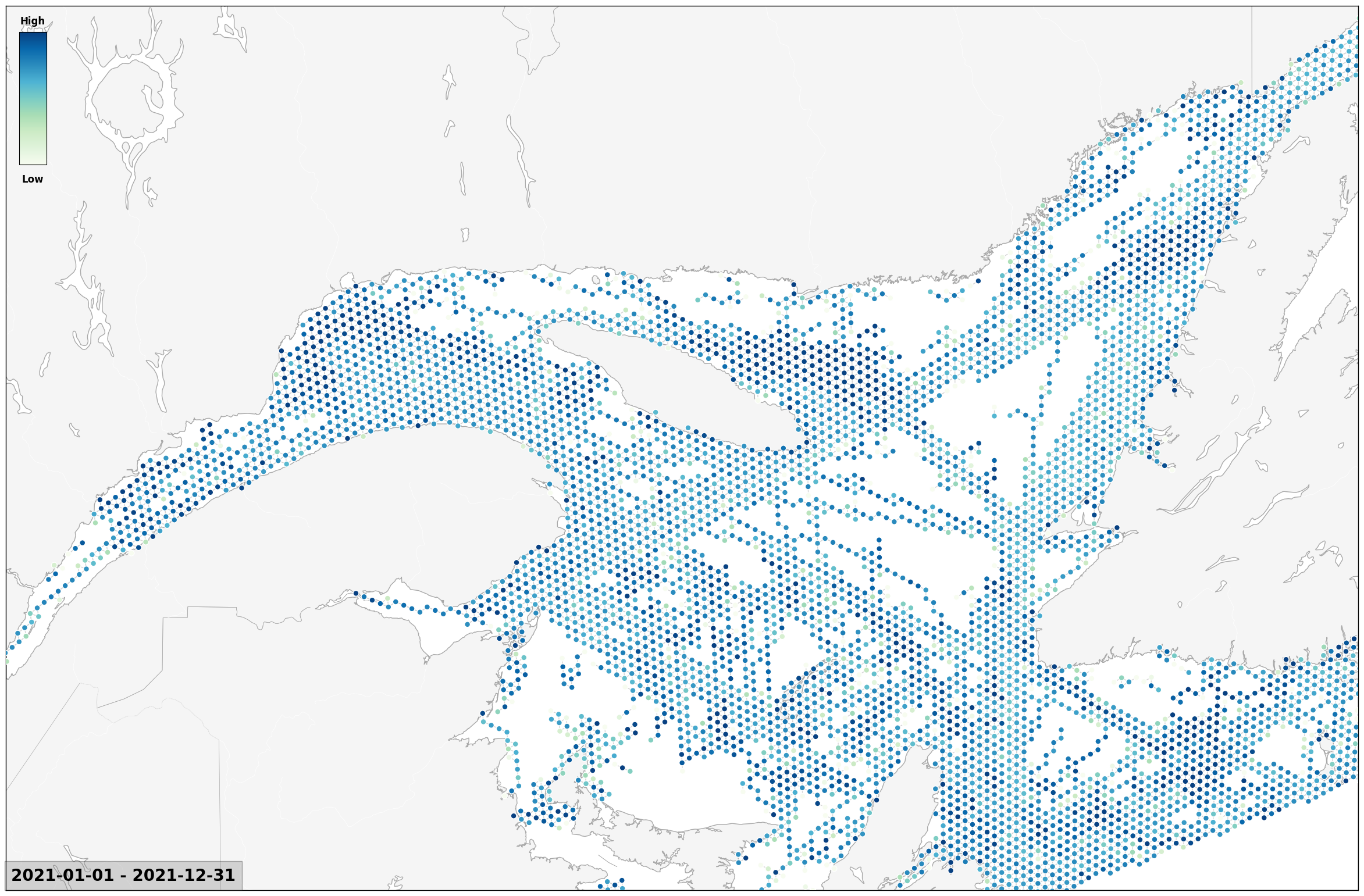} &
        \includegraphics[width=0.25\linewidth]{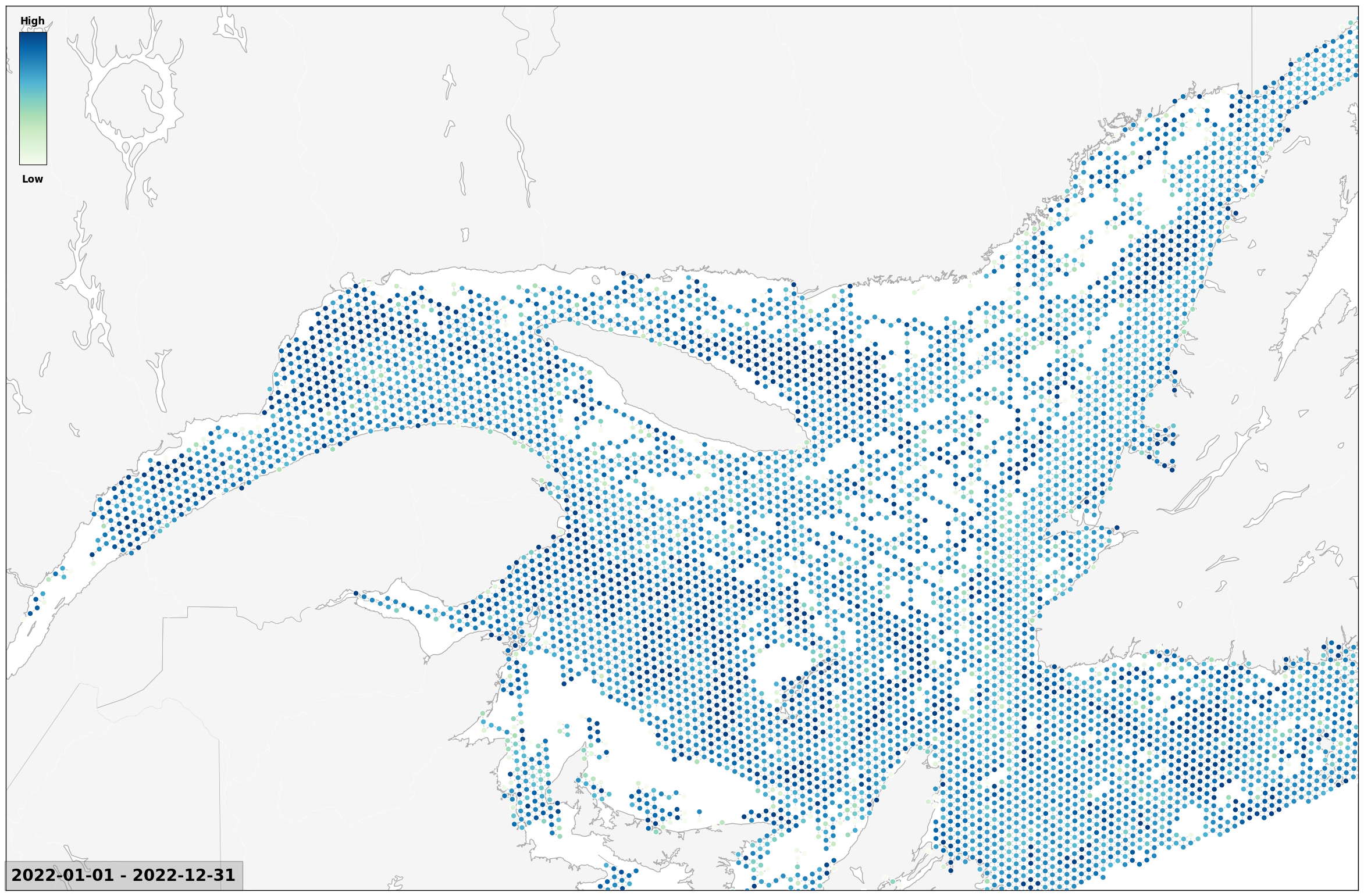} \\
    \end{tabular}
    \caption{Average dwell of multiple ships per cell patterns from 2019 to 2022 for fishing vessels.}
    \label{fig:fishing_dwell}
\end{figure*}

Figure~\ref{fig:fishing_dwell} presents the spatial distribution of \textit{mean dwell time} for fishing vessels, defined as the average dwell per cell aggregated across all vessels within a given year. In 2019, only approximately 10\% of the hexagonal grid recorded non-zero dwell times, forming distinct clusters primarily along the Lower North Shore, around Anticosti Island, and in Gaspé. 

During the pandemic year 2020, the spatial extent of these dwell regions expanded significantly. The proportion of active cells increased by roughly 31\%, with new clusters emerging eastward along the Laurentian Channel and extending southward toward the Cabot Strait. This expansion continued into 2021 when the footprint reached its peak at nearly 15\% of the grid (46\% higher relative to the 2019 baseline) before stabilizing in 2022 at approximately 35\% above the baseline.

\begin{figure*}[b]
    \centering\hspace{-.95cm}
    \setlength{\tabcolsep}{1pt}
    \renewcommand{\arraystretch}{0.9}
    \begin{tabular}{@{}r@{\hspace{4pt}}c@{\hspace{2pt}}c@{\hspace{2pt}}c@{\hspace{2pt}}c@{}}
        & \multicolumn{1}{c}{\textit{\textbf{All}}} &
          \multicolumn{1}{c}{\textit{\textbf{Commercial}}} &
          \multicolumn{1}{c}{\textit{\textbf{Fishing}}} &
          \multicolumn{1}{c}{\textit{\textbf{Passenger}}} \\[1.5pt]
        \raisebox{1\height}{\rotatebox[origin=c]{90}{\hspace{.5cm}\textit{\textbf{Transition}}}} &
        \includegraphics[width=0.25\linewidth]{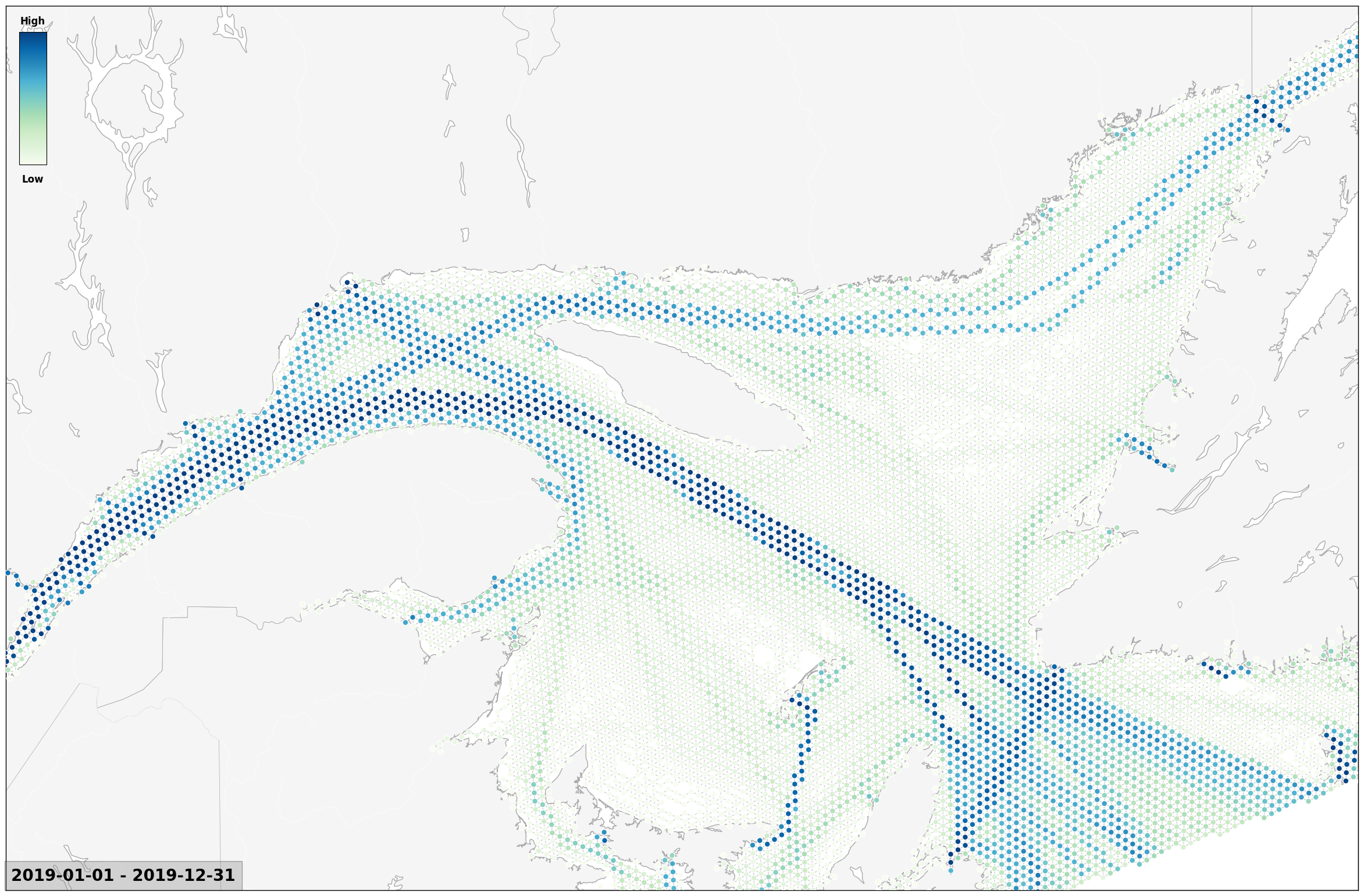} &
        \includegraphics[width=0.25\linewidth]{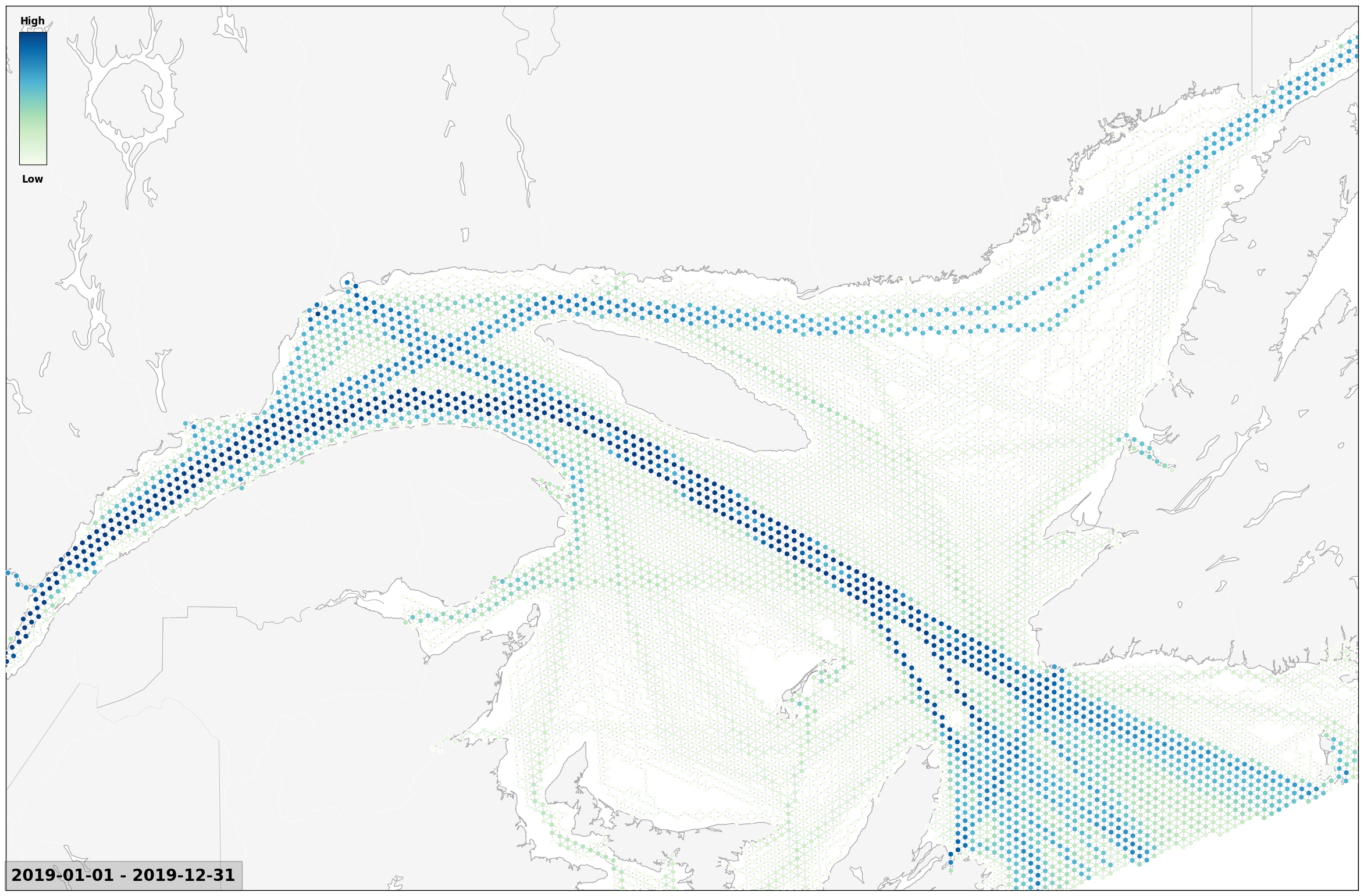} &
        \includegraphics[width=0.25\linewidth]{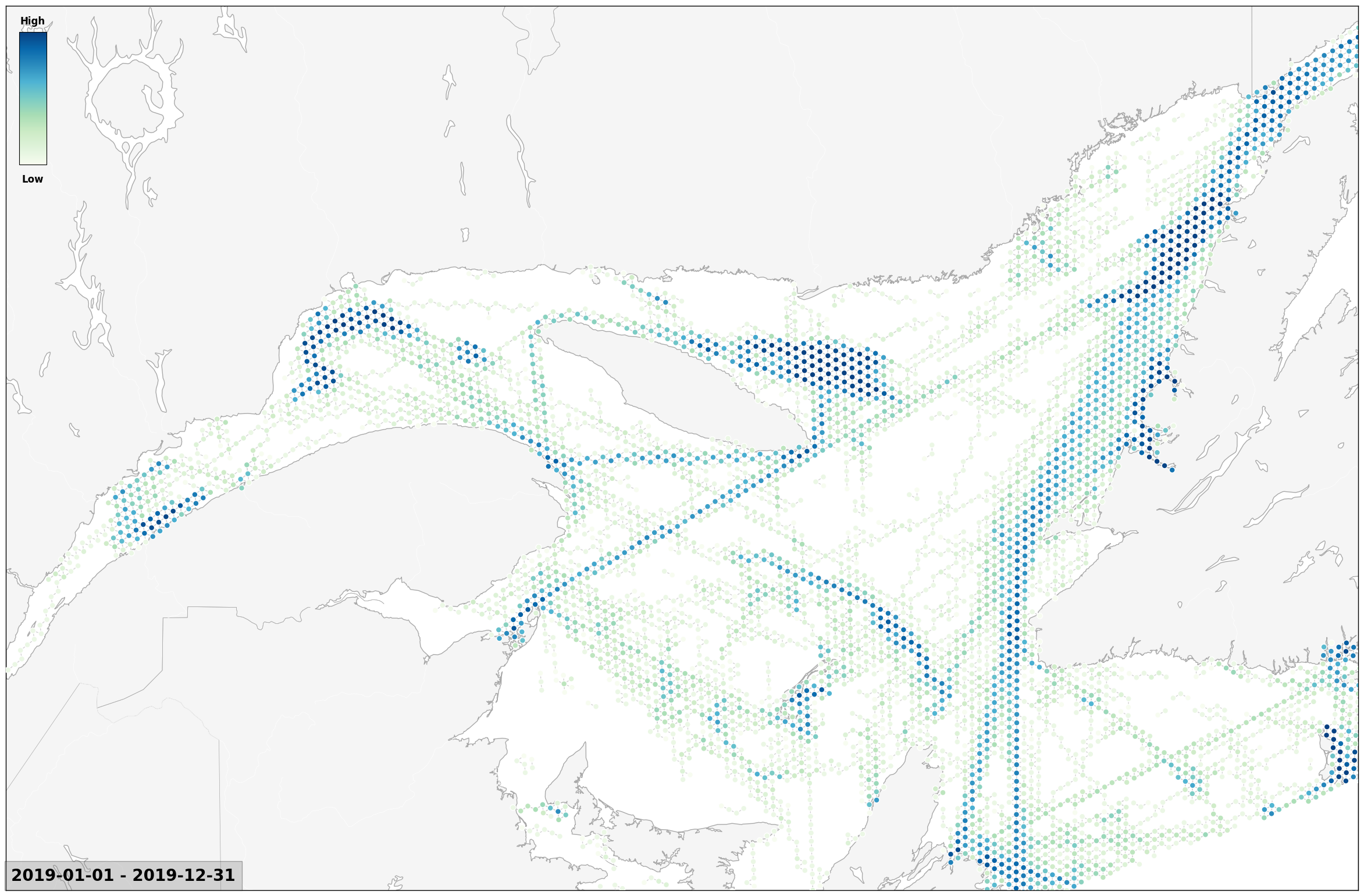} &
        \includegraphics[width=0.25\linewidth]{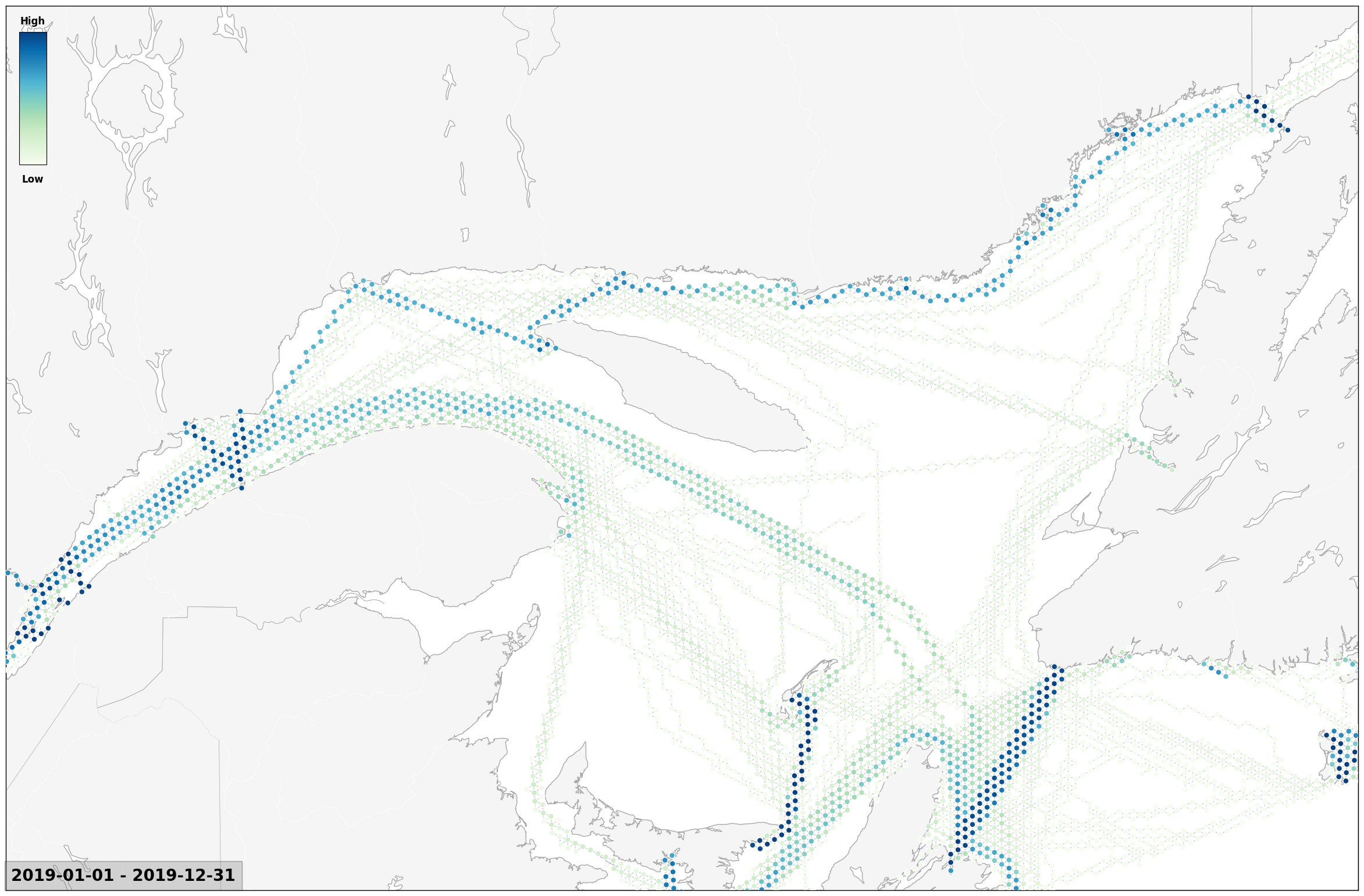} \\[1pt]
        \raisebox{1\height}{\rotatebox[origin=c]{90}{\hspace{1cm}\textit{\textbf{Dwell}}}} &
        \includegraphics[width=0.25\linewidth]{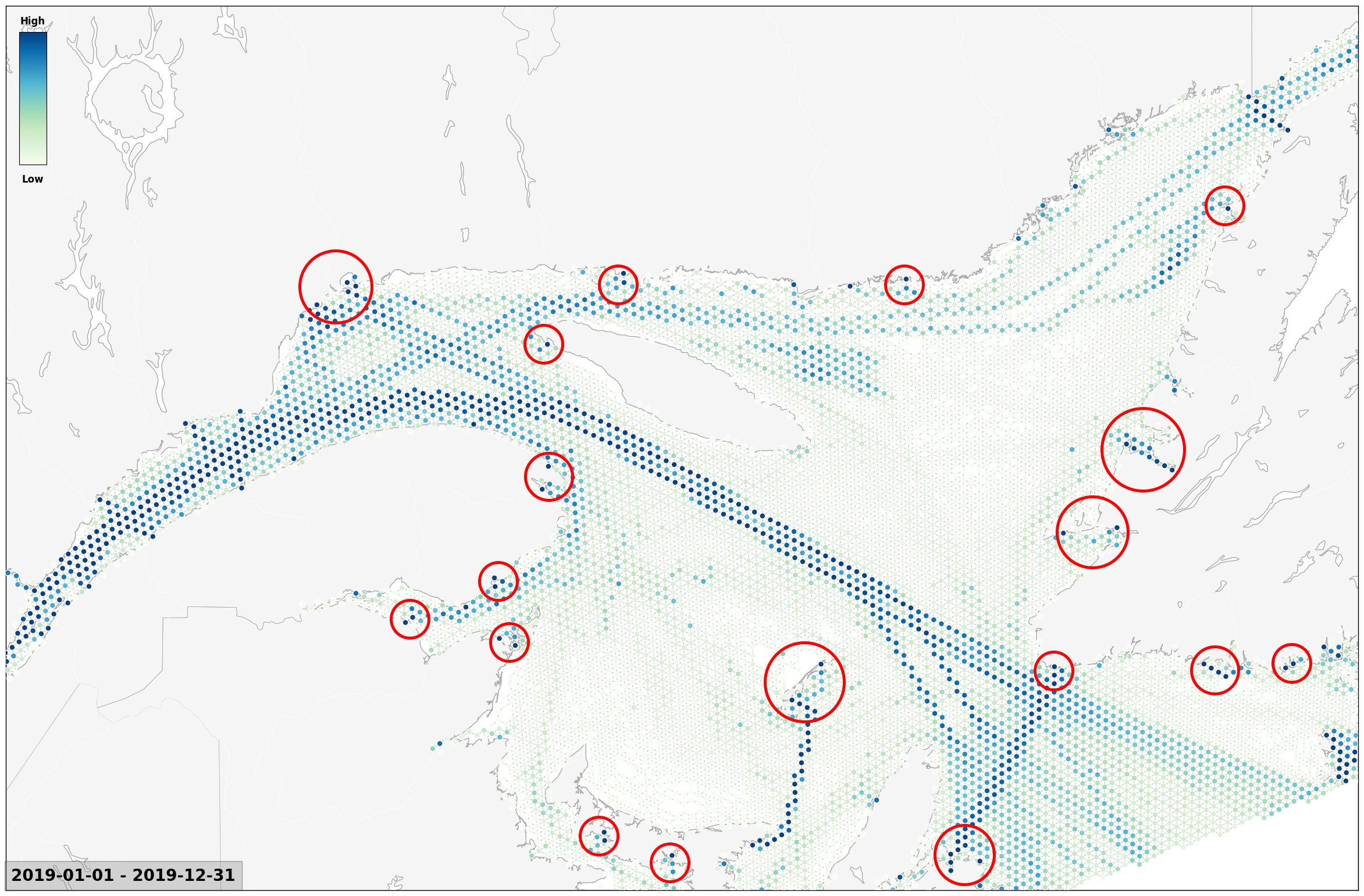} &
        \includegraphics[width=0.25\linewidth]{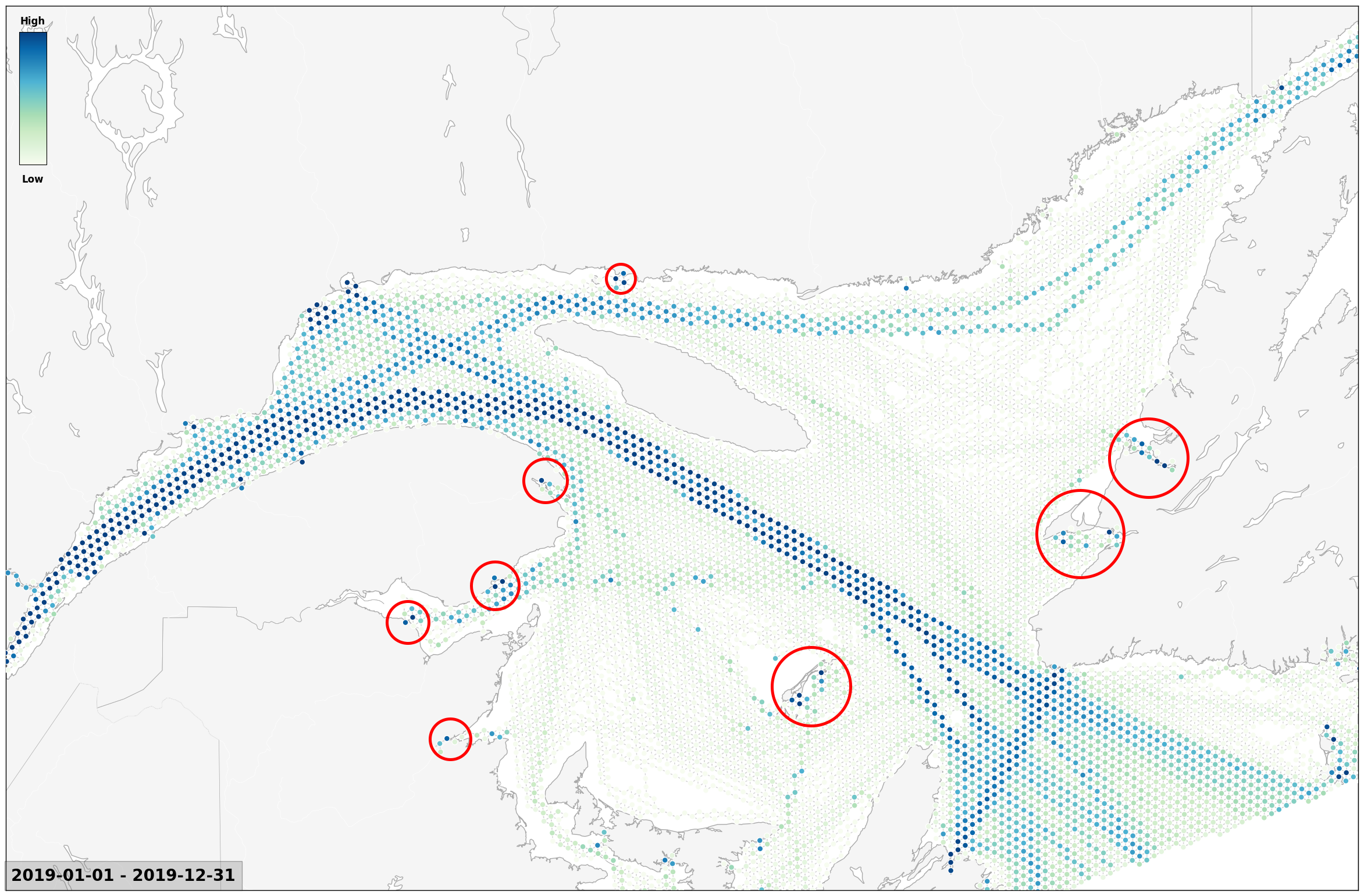} &
        \includegraphics[width=0.25\linewidth]{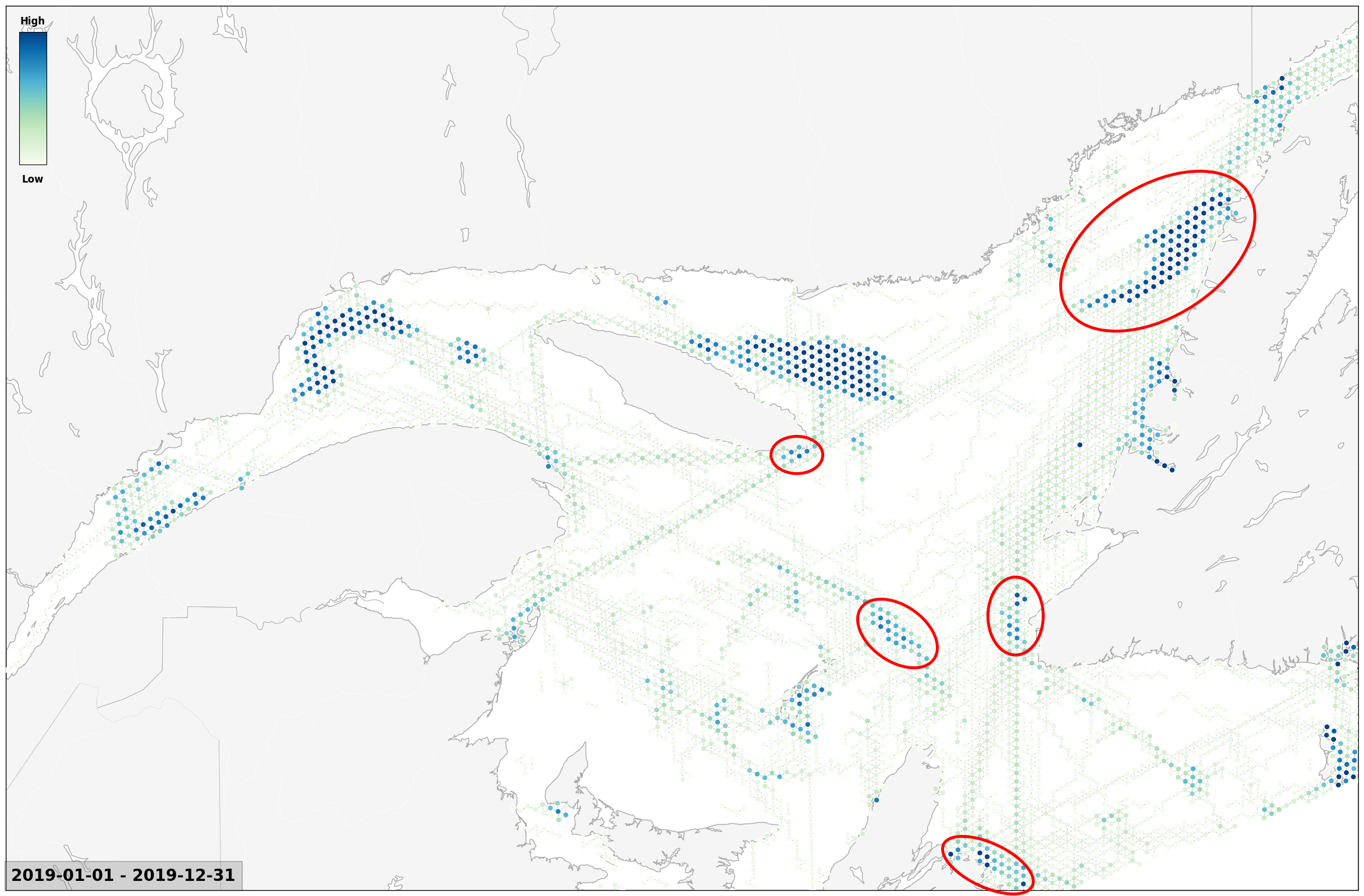} &
        \includegraphics[width=0.25\linewidth]{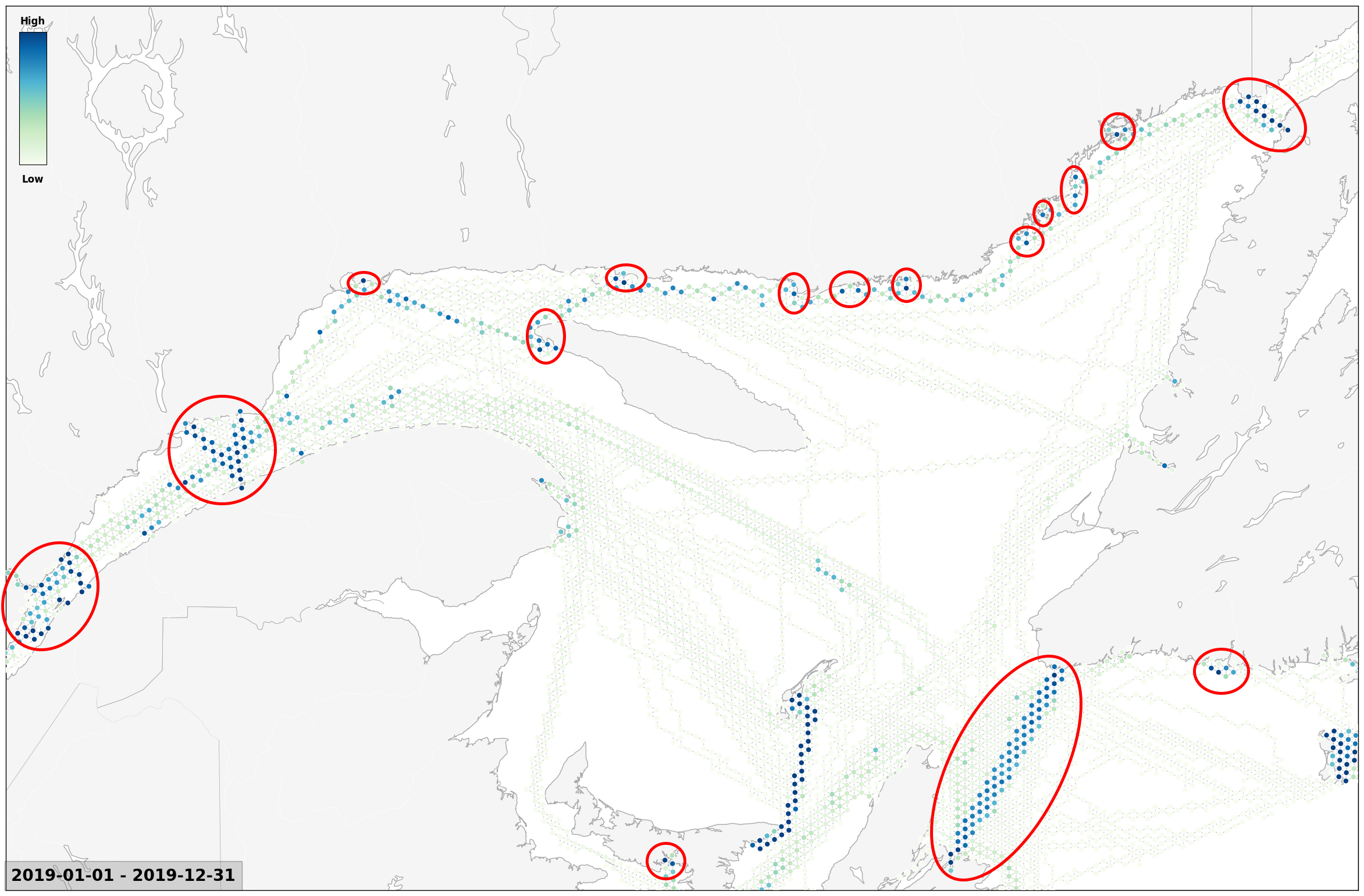} \\
    \end{tabular}
    \caption{Transition and dwell patterns for 2019 (accumulated), by vessel-type category.}
    \label{fig:transition_dwell_2019}
\end{figure*}

In parallel with the spatial enlargement, the dwell heatmap becomes increasingly contiguous over time. By 2022, the number of disconnected patches visibly declines, indicating that previously isolated hotspots have become connected through low- to moderate-dwell cells. This spatial mosaic, coupled with the plateauing of system-level dwell magnitude, suggests a behavioral transition from concentrated, site-specific activity to a more spatially distributed yet temporally consistent operational pattern. Practically, fishing vessels appear to have adapted to post-pandemic conditions by expanding their grounds and smoothing dwell durations across a broader portion of the Gulf, rather than reverting to focal zones.

\vspace{.1cm}
\noindent\textit{\textbf{Variations in Spatial Signatures by Vessel Type}}
\vspace{.1cm}

Figure~\ref{fig:transition_dwell_2019} provides a spatially disaggregated overview of vessel mobility patterns during 2019 (the baseline period, $\mathcal{T}_{\text{pre}}$), stratified by vessel type. The upper row shows cumulative transition counts, while the lower row displays total dwell-time intensity across the discretized maritime space. Each subplot corresponds to a distinct vessel category, allowing for a direct comparison of characteristic mobility signatures.

In the transition panels (top row), the dominant maritime corridor structure is characterized by dense east-west flows that trace the main shipping lanes of the Gulf of St. Lawrence. The \textit{All} and \textit{Commercial} vessel panels exhibit similar high-density tracks along the \textit{Laurentian Channel}, indicating that commercial vessels largely define the region's core traffic structure. The \textit{Fishing} panel, in contrast, reveals more dispersed and transversal trajectories, frequently branching perpendicularly from major corridors into shelf and coastal zones. The \textit{Passenger} layer shows highly localized transitions between coastal terminals and island settlements.

The dwell panels (bottom row) highlight spatial zones where vessels remain stationary for extended periods. For the \textit{Commercial} group, prolonged dwell is concentrated near major ports and anchorages ({\it e.g.}, Québec City, Sept-Îles, and the approaches to Halifax), forming distinct clusters. \textit{Fishing} vessels exhibit elevated dwell in shallow shelf areas, particularly around Anticosti Island and the Gaspé Peninsula, which correspond to established fishing grounds and indicate persistent operational activity~\cite{gulffishing2021}. For the \textit{Passenger} category, dwell time is concentrated at terminal endpoints and along the Lower North Shore, consistent with fixed-route ferry services and seasonal cruise stopovers.

\begin{figure*}[h]
    \centering\hspace{-.75cm}
    \setlength{\tabcolsep}{2pt} 
    \renewcommand{\arraystretch}{0.9} 
    \begin{tabular}{@{}r@{\hspace{4pt}}c@{\hspace{2pt}}c@{\hspace{2pt}}c@{}}
        & \multicolumn{1}{c}{\textit{\textbf{2020}}} & 
          \multicolumn{1}{c}{\textit{\textbf{2021}}} & 
          \multicolumn{1}{c}{\textit{\textbf{2022}}} \\[-1.5pt]
        \raisebox{1\height}{\rotatebox[origin=c]{90}{\hspace{0.8cm}\textit{\textbf{Commercial}}}} &
        \includegraphics[width=0.33\linewidth]{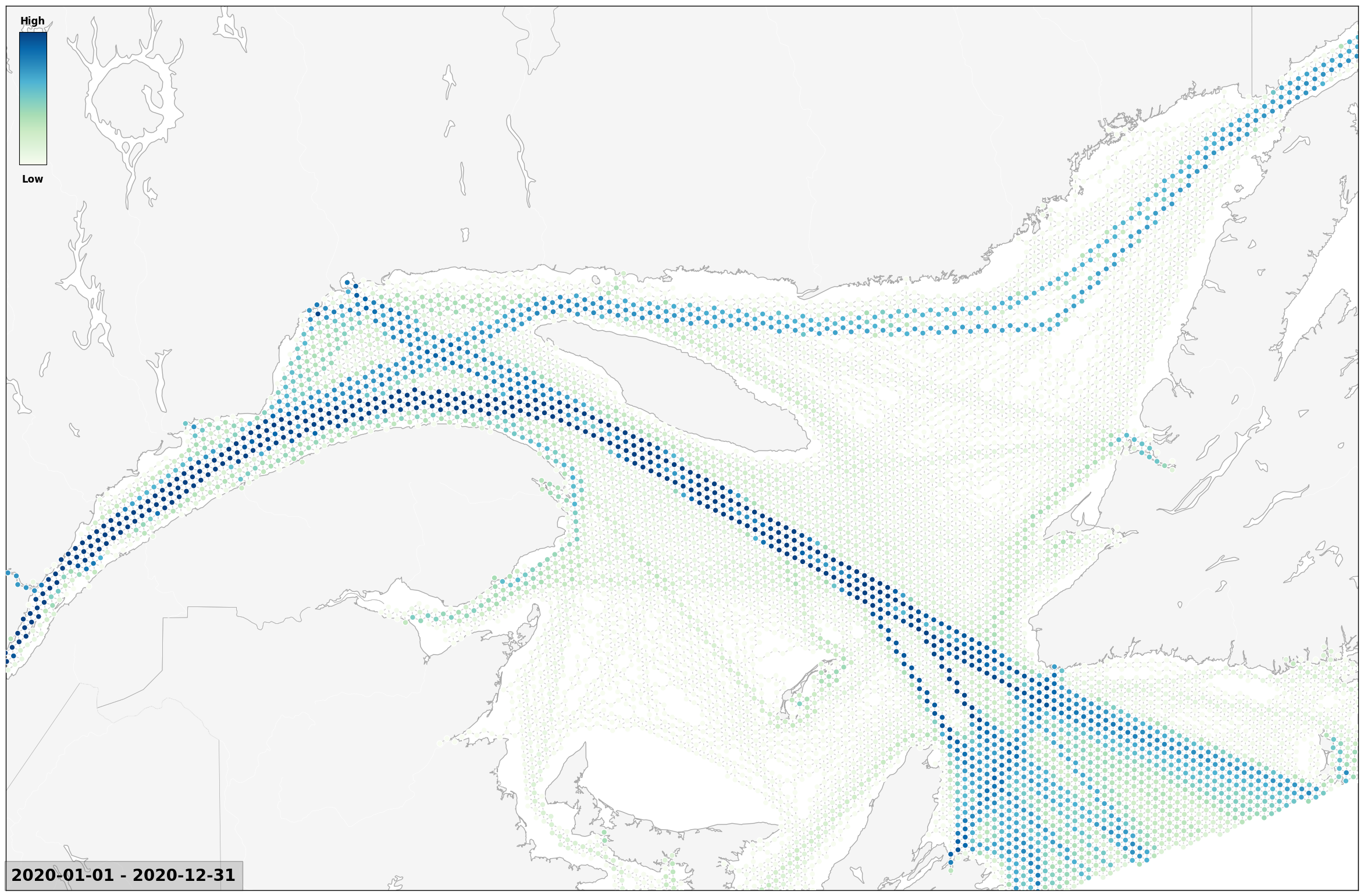} &
        \includegraphics[width=0.33\linewidth]{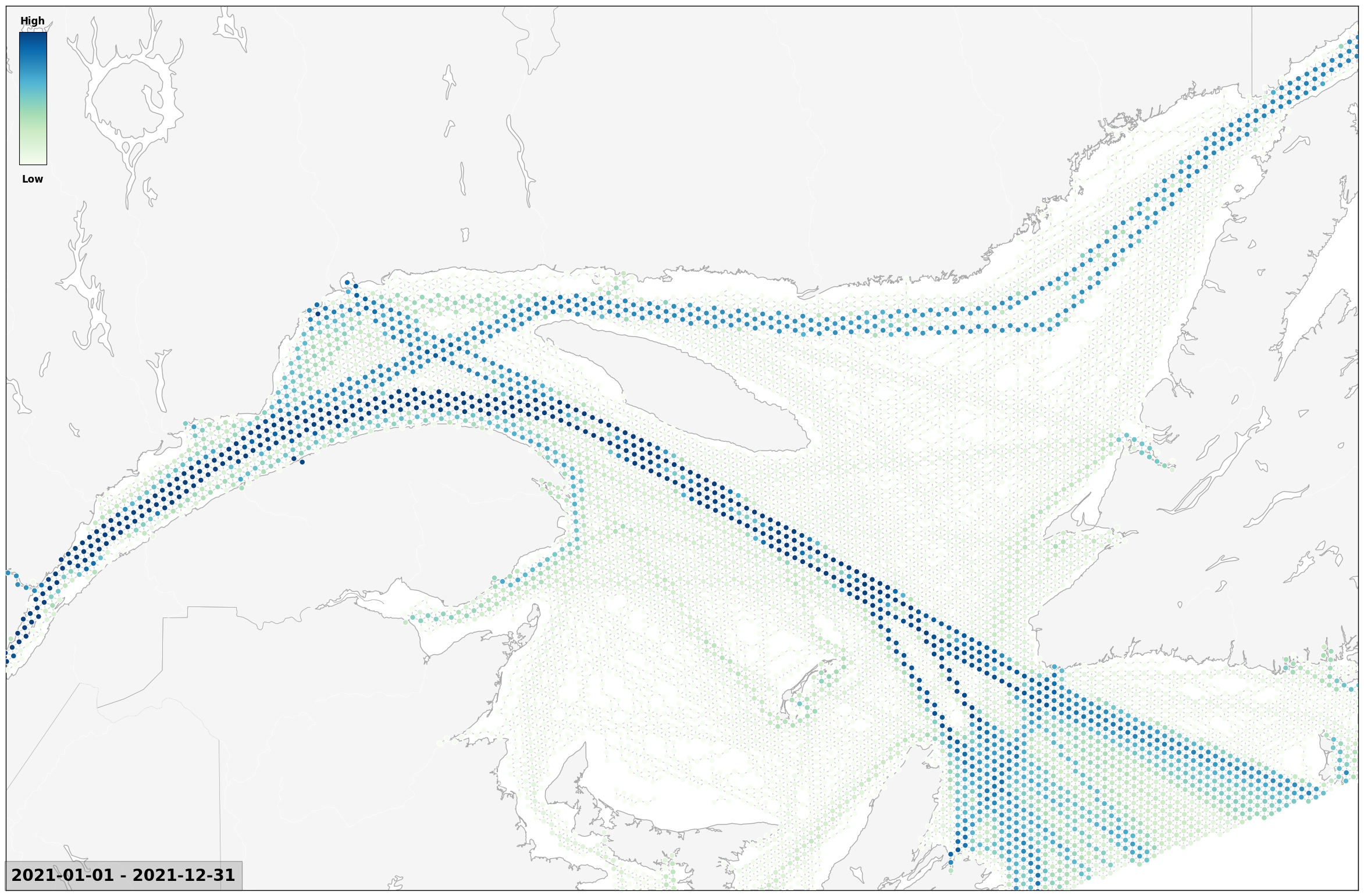} &
        \includegraphics[width=0.33\linewidth]{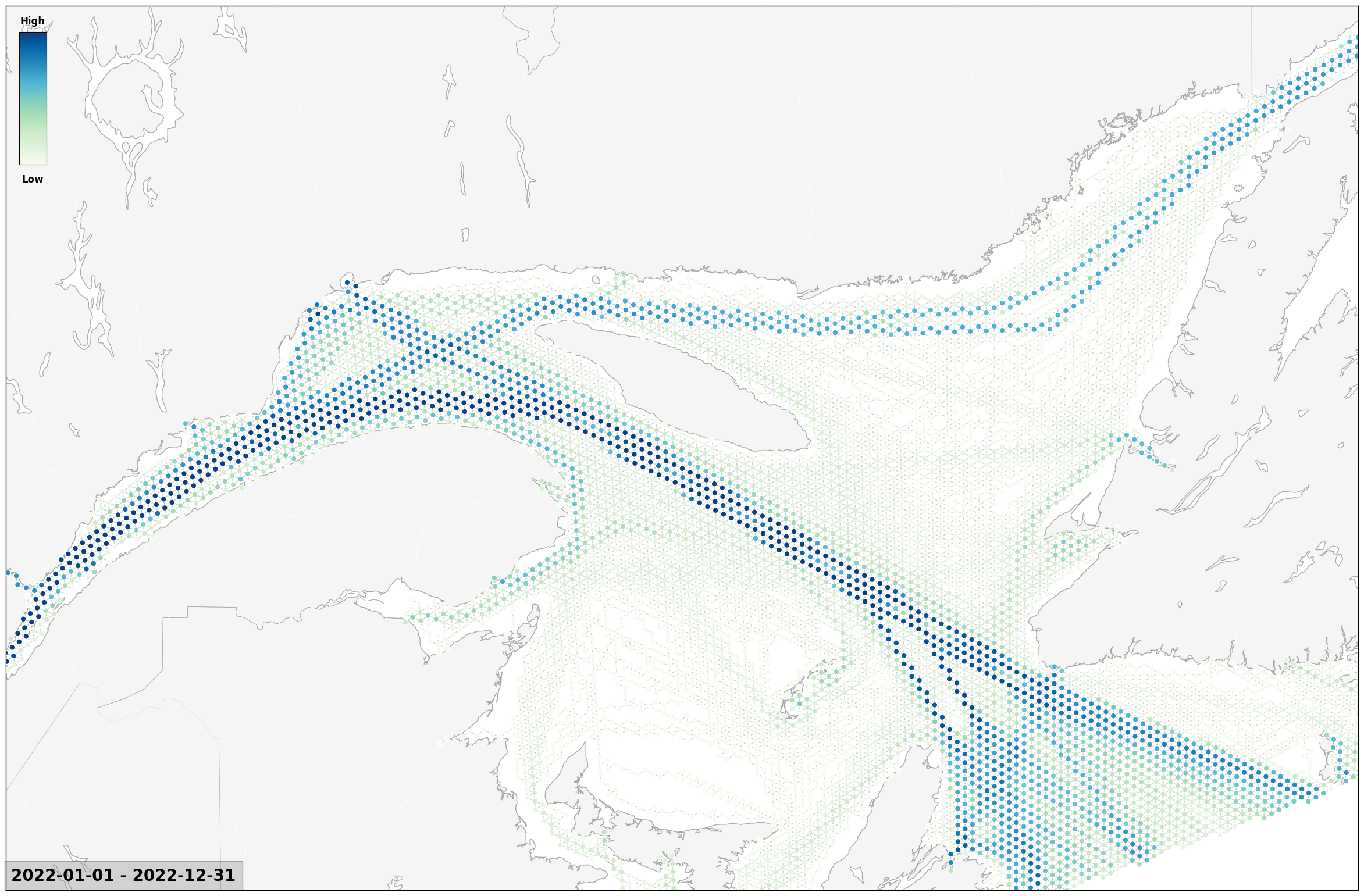} \\
        \raisebox{1\height}{\rotatebox[origin=c]{90}{\hspace{1.3cm}\textit{\textbf{Fishing}}}} &
        \includegraphics[width=0.33\linewidth]{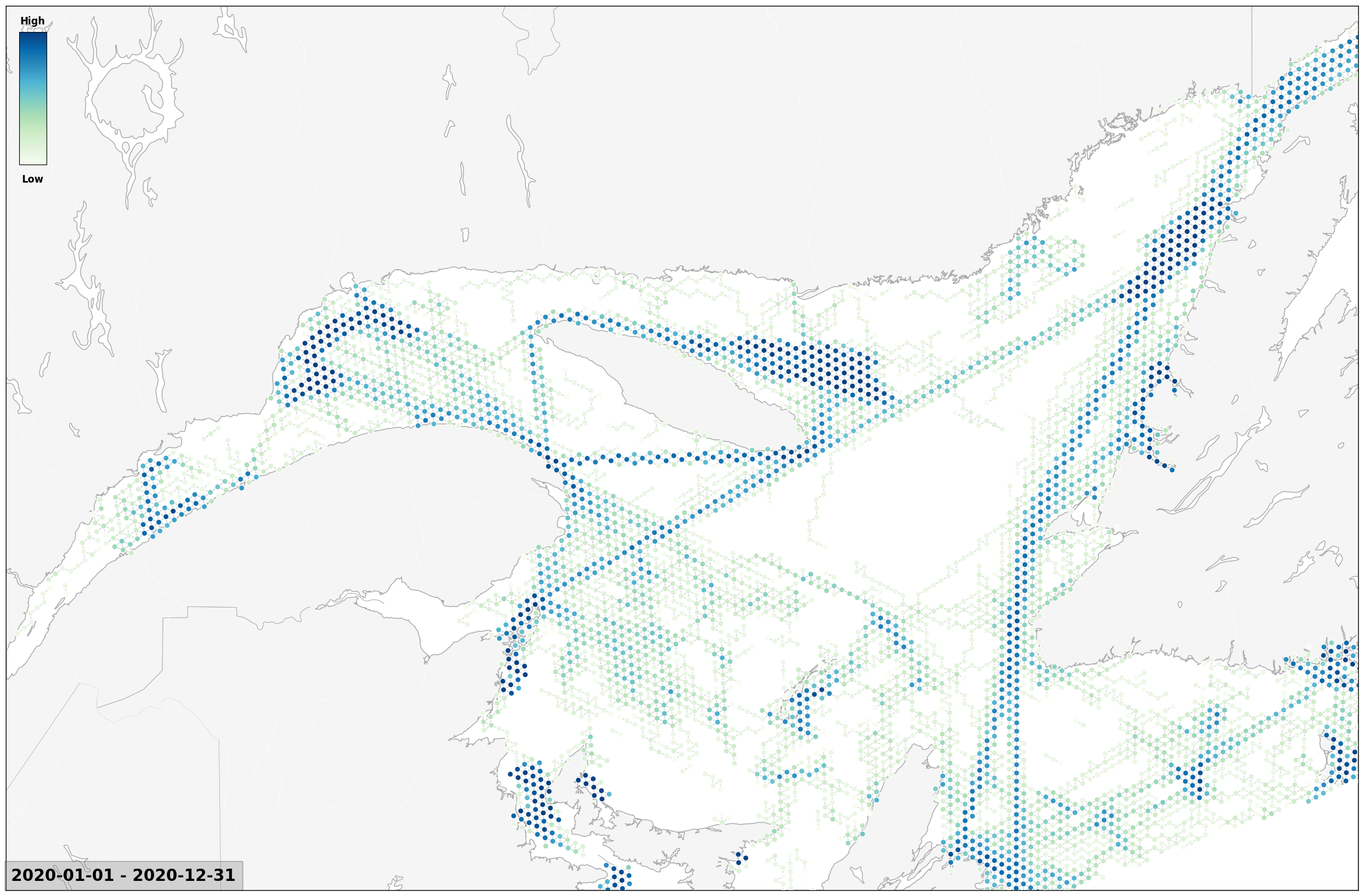} &
        \includegraphics[width=0.33\linewidth]{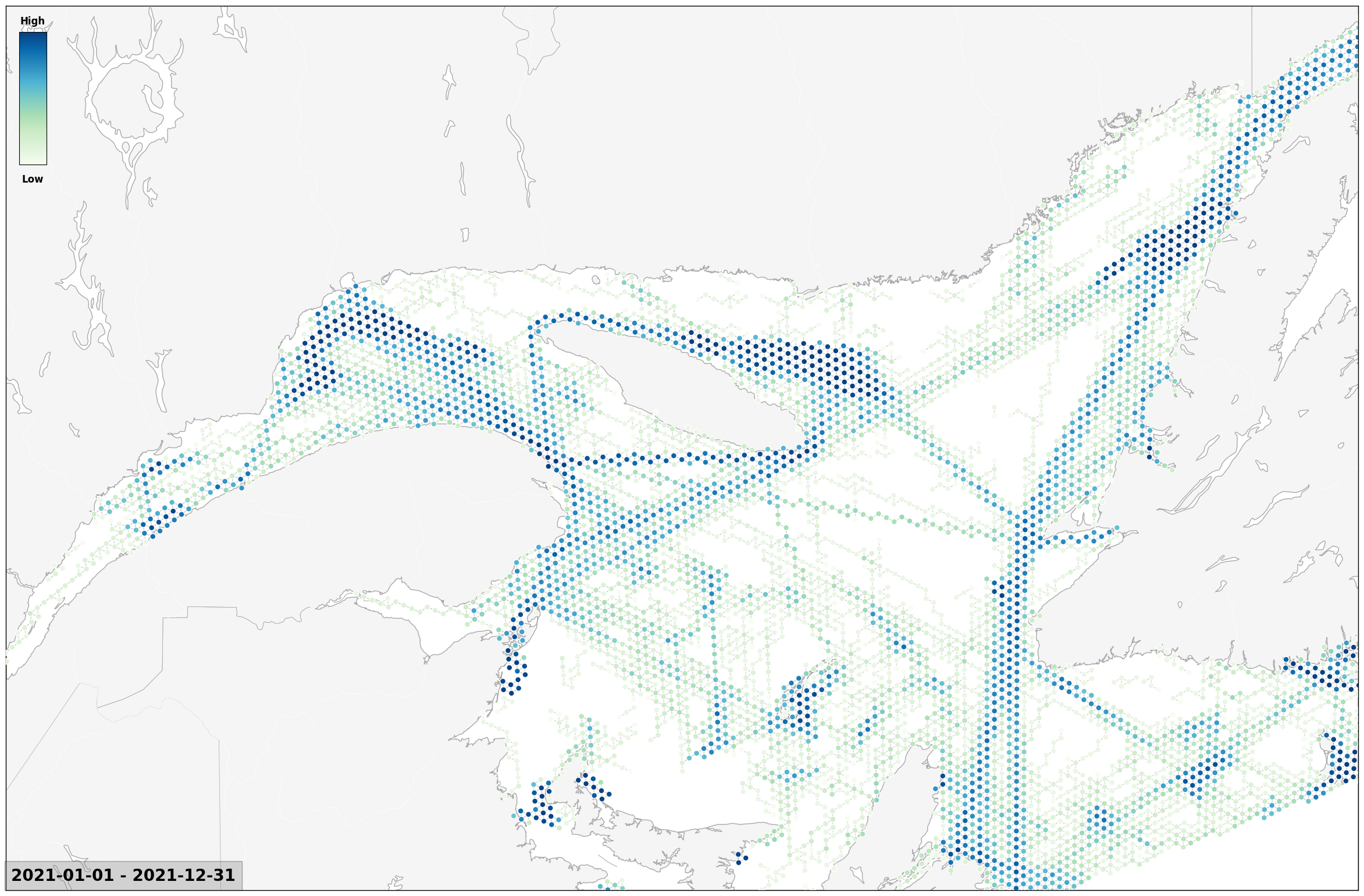} &
        \includegraphics[width=0.33\linewidth]{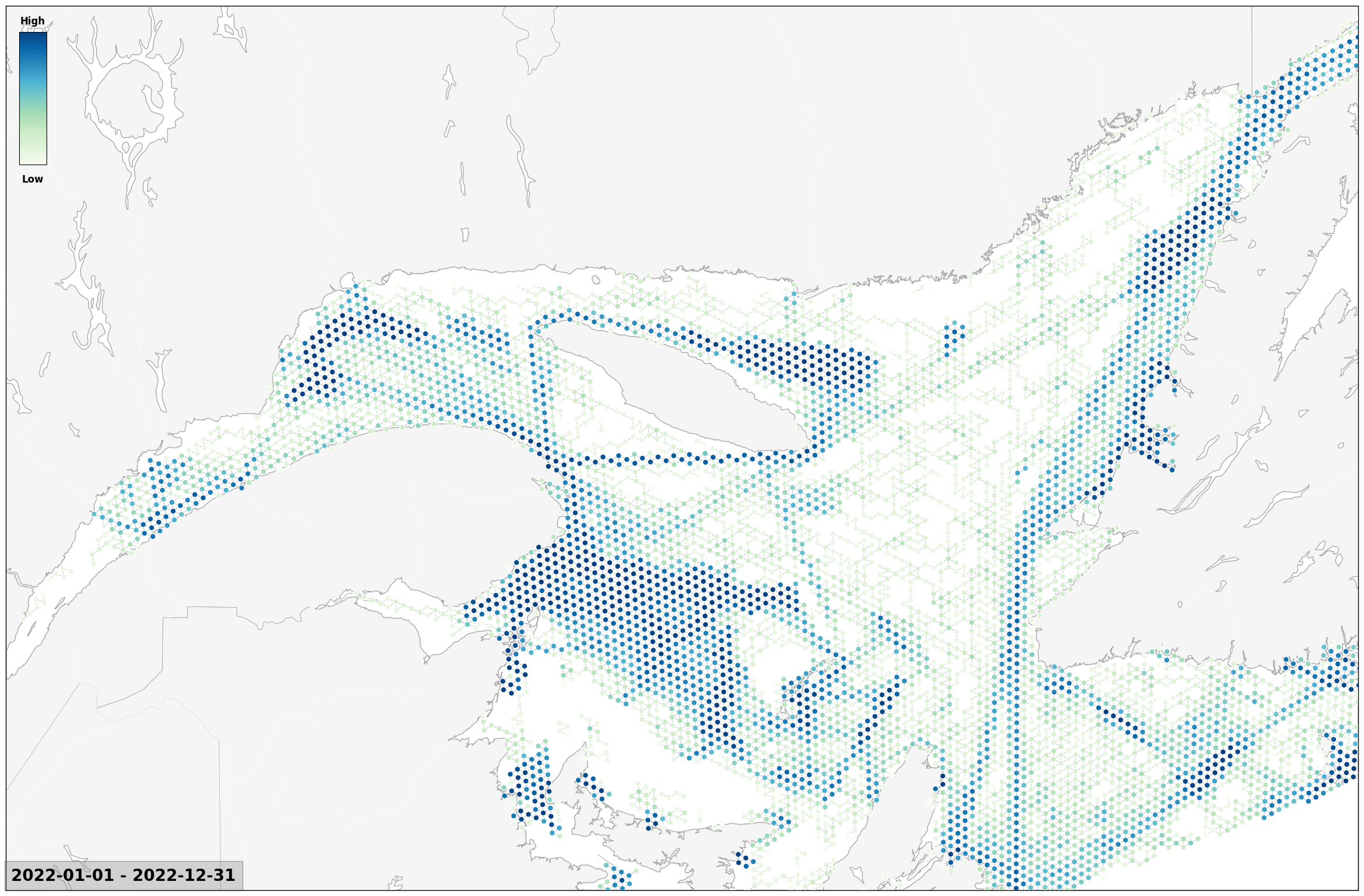} \\
        \raisebox{1\height}{\rotatebox[origin=c]{90}{\hspace{1.1cm}\textit{\textbf{Passenger}}}} &
        \includegraphics[width=0.33\linewidth]{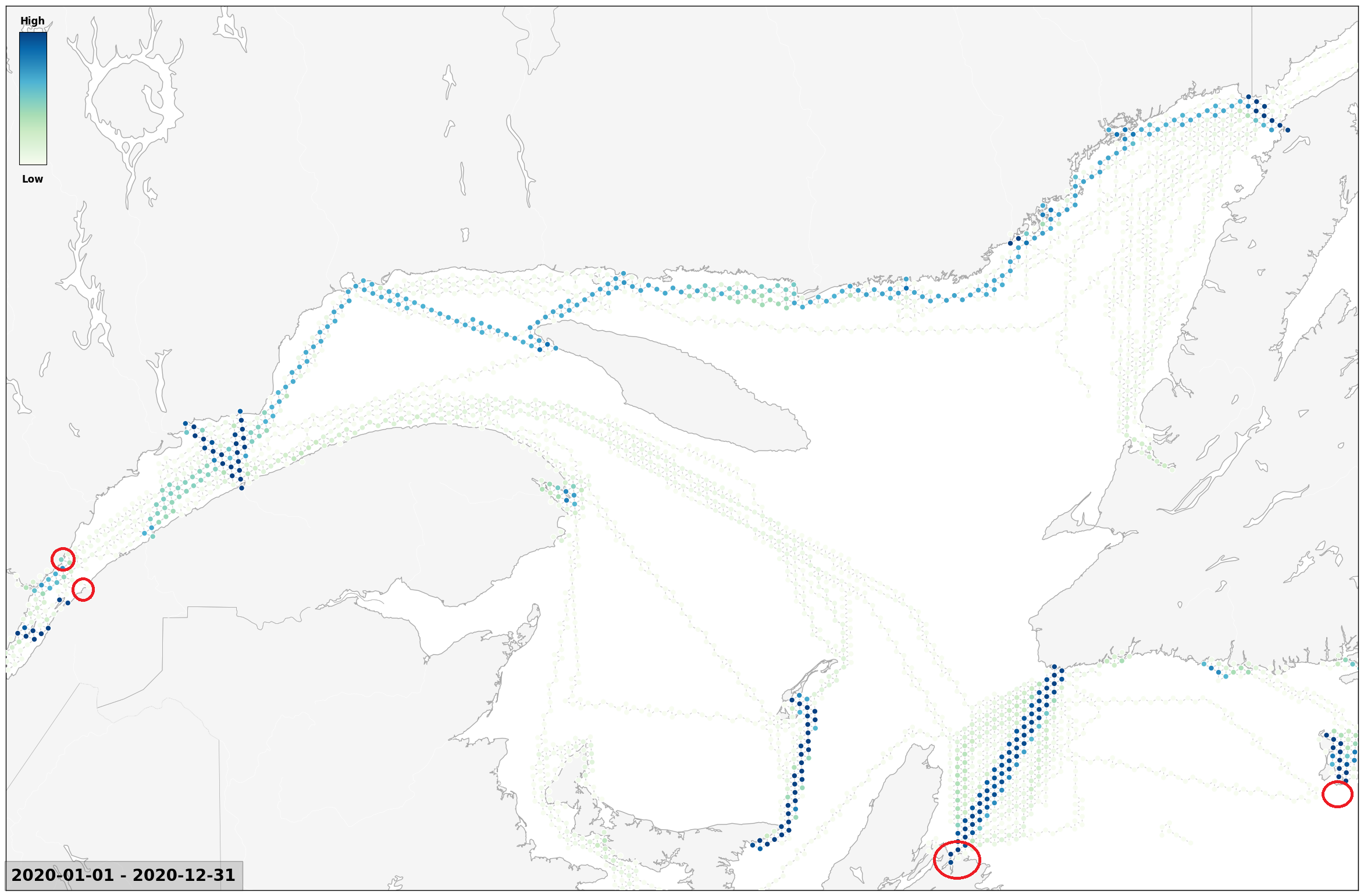} &
        \includegraphics[width=0.33\linewidth]{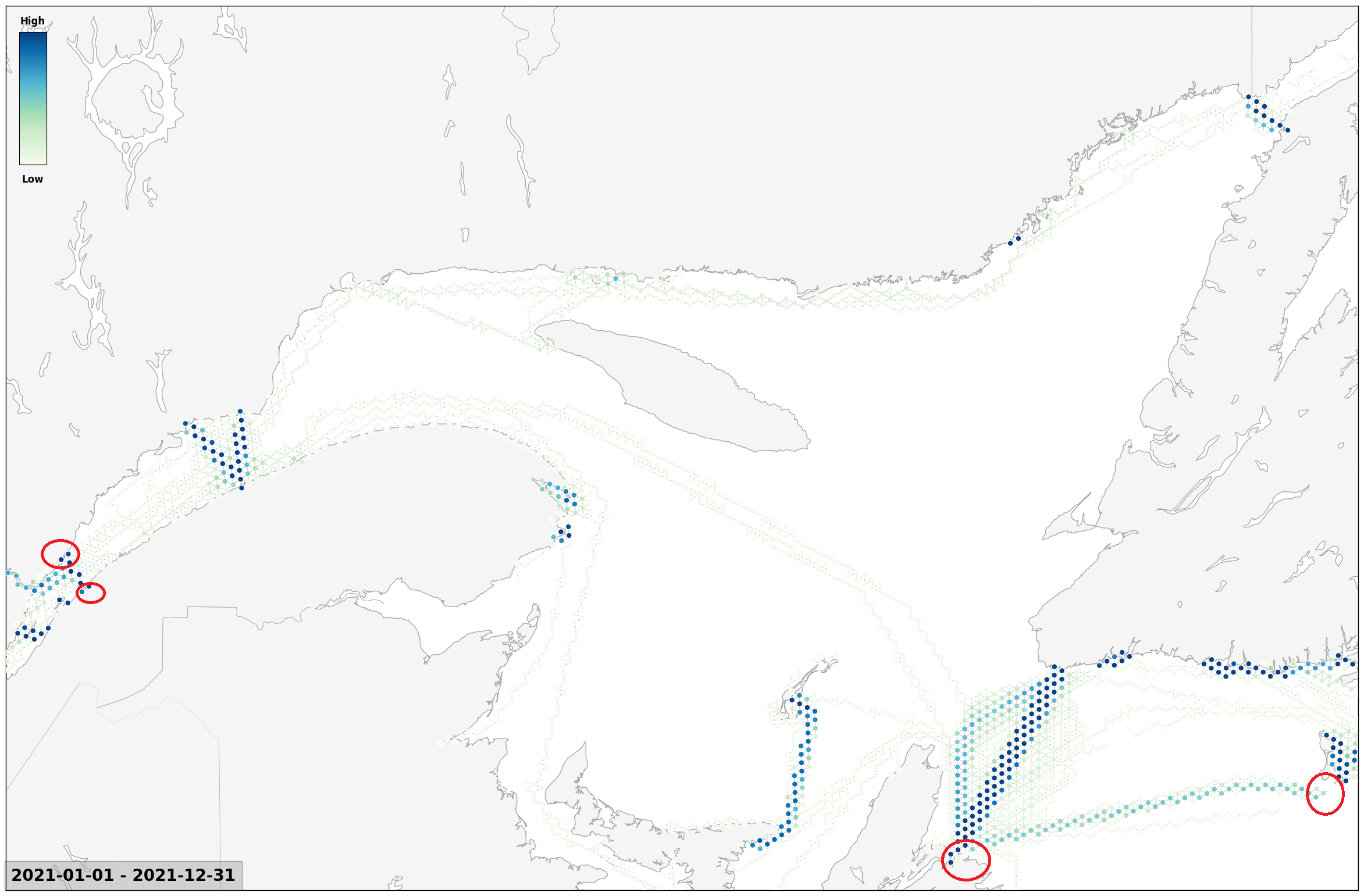} &
        \includegraphics[width=0.33\linewidth]{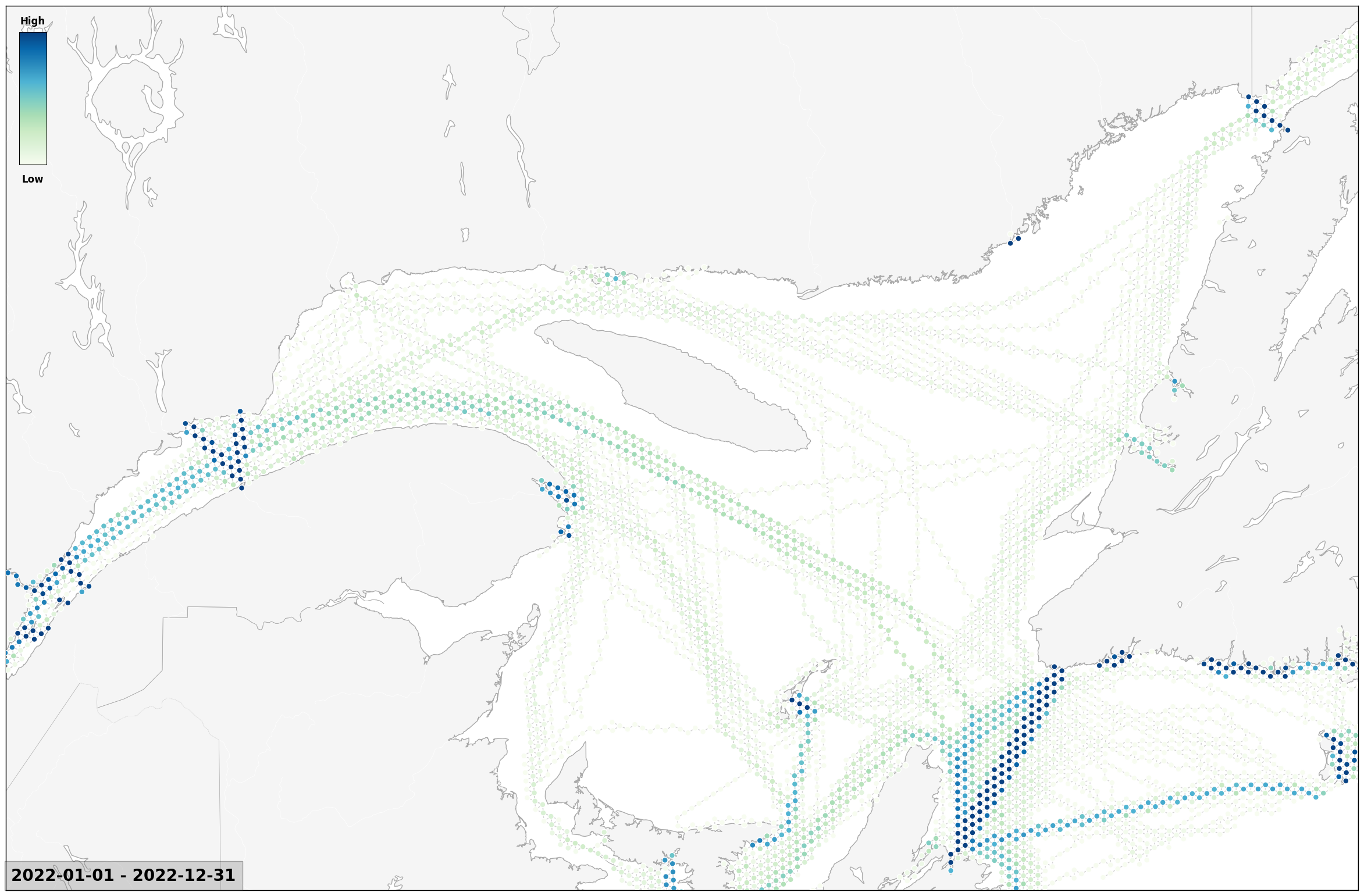} \\
    \end{tabular}
    \caption{\textbf{Raw Transition Intensity.} Spatial footprint of cell–visit counts by vessel type across 2020--2022.}
    \label{fig:transition}
\end{figure*}

\begin{figure*}[ht]
    \centering\hspace{-.75cm}
    \setlength{\tabcolsep}{2pt} 
    \renewcommand{\arraystretch}{0.9} 
    \begin{tabular}{@{}r@{\hspace{4pt}}c@{\hspace{2pt}}c@{\hspace{2pt}}c@{}}
          & \multicolumn{1}{c}{\textit{\textbf{2020}}} &
          \multicolumn{1}{c}{\textit{\textbf{2021}}} & 
          \multicolumn{1}{c}{\textit{\textbf{2022}}} \\[-1.5pt]
        \raisebox{1\height}{\rotatebox[origin=c]{90}{\hspace{0.8cm}\textit{\textbf{Commercial}}}} &
        \includegraphics[width=0.33\linewidth]{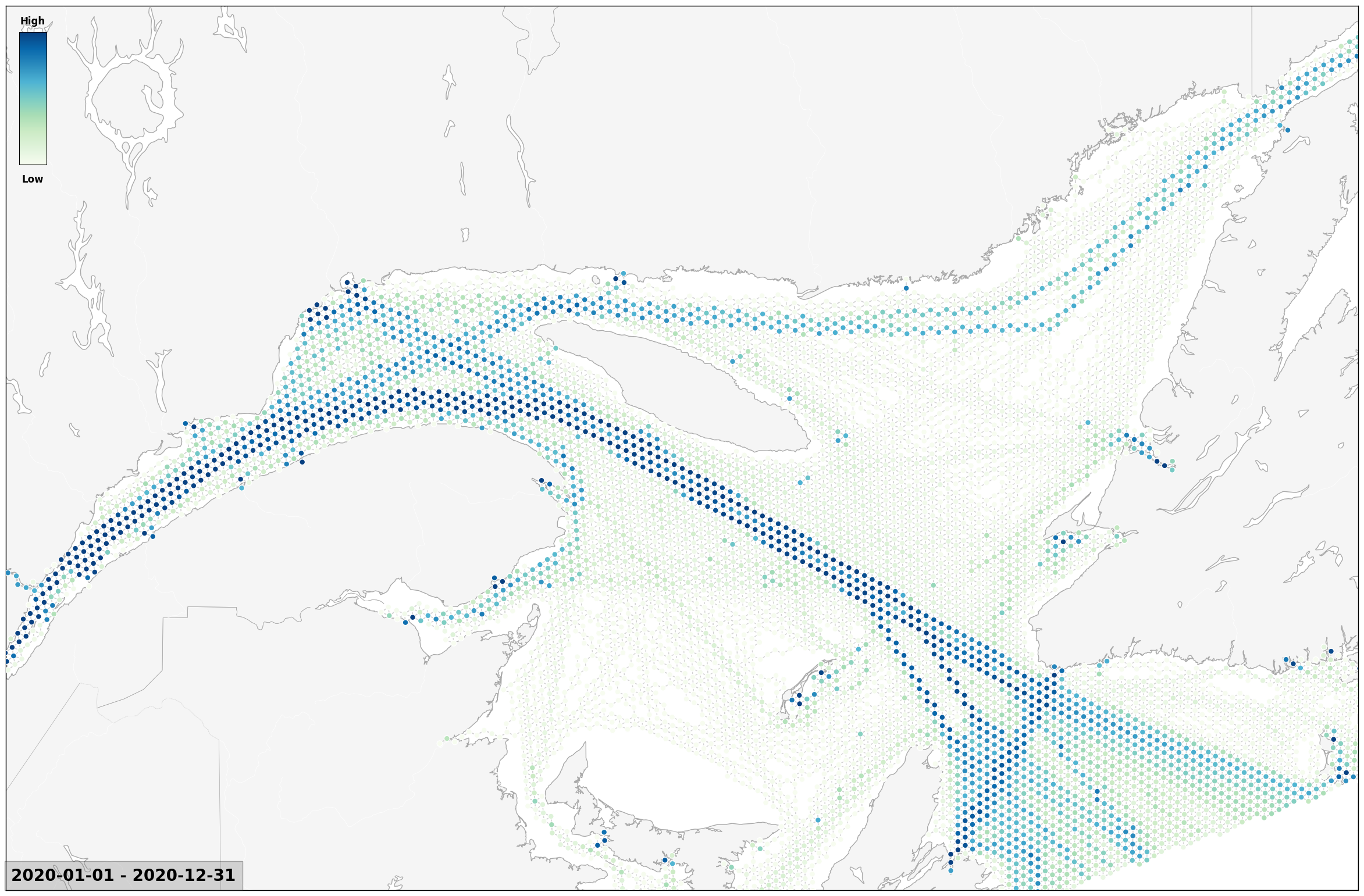} &
        \includegraphics[width=0.33\linewidth]{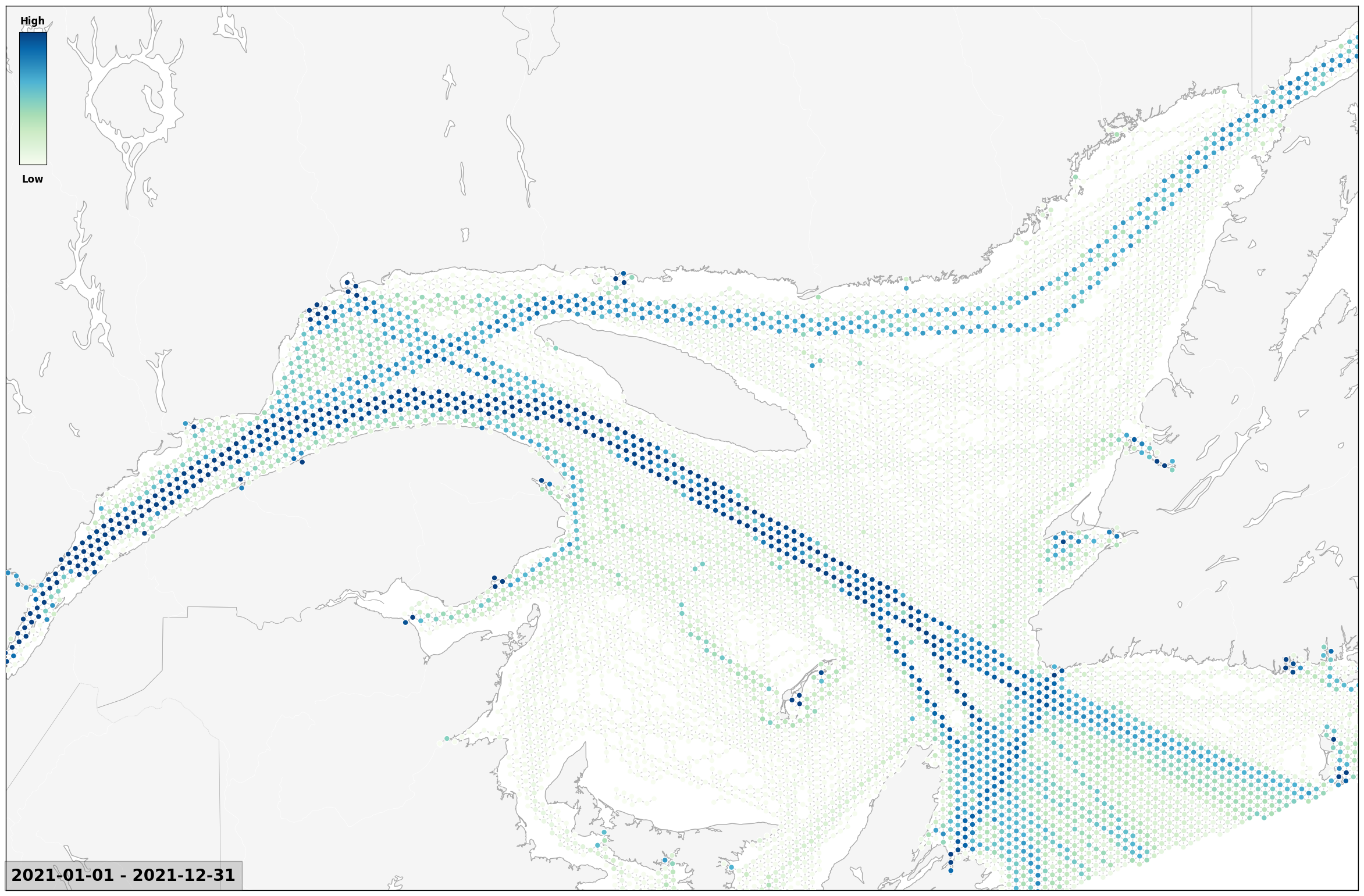} &
        \includegraphics[width=0.33\linewidth]{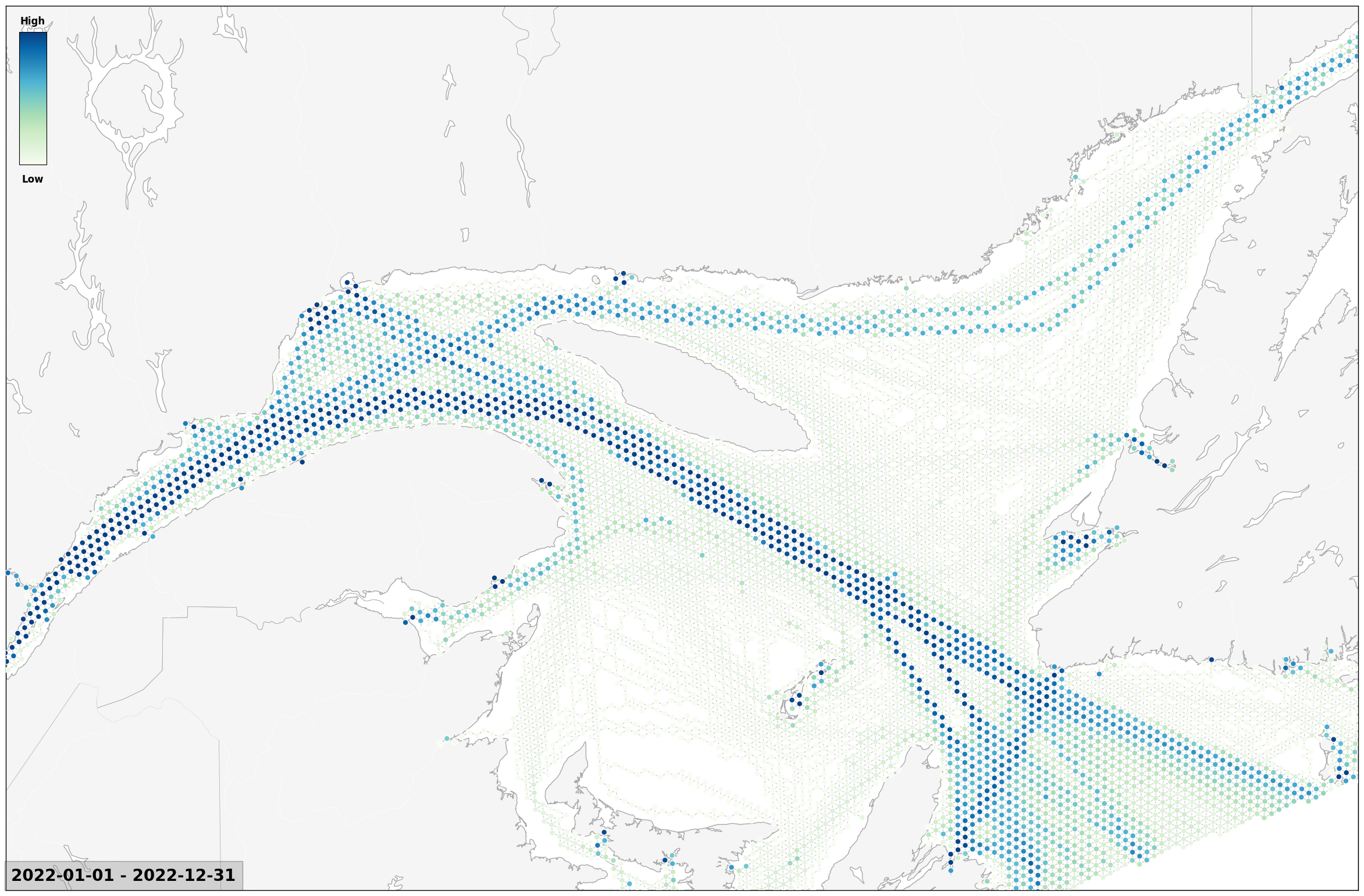} \\
        \raisebox{1\height}{\rotatebox[origin=c]{90}{\hspace{1.3cm}\textit{\textbf{Fishing}}}} &
        \includegraphics[width=0.33\linewidth]{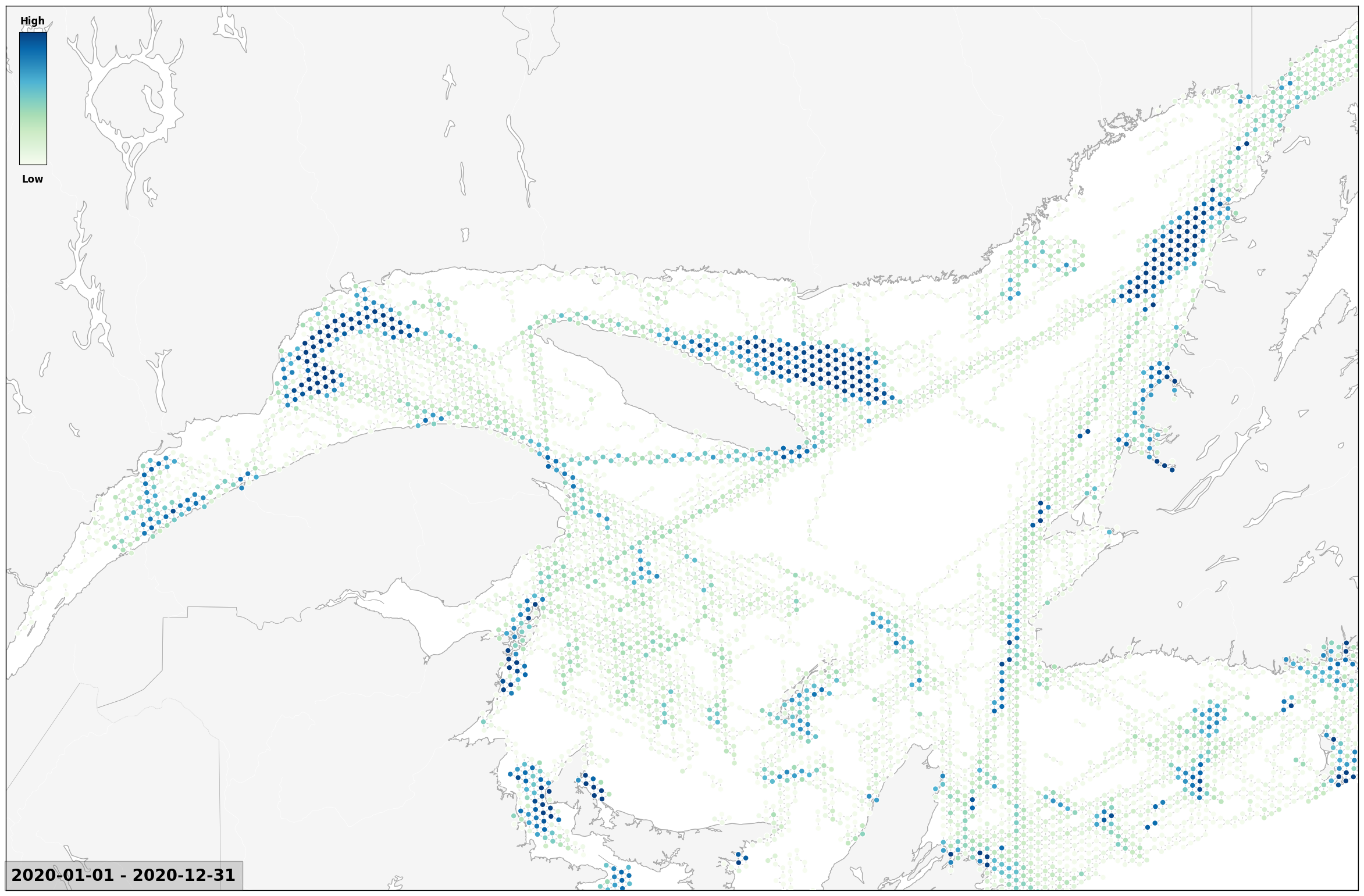} &
        \includegraphics[width=0.33\linewidth]{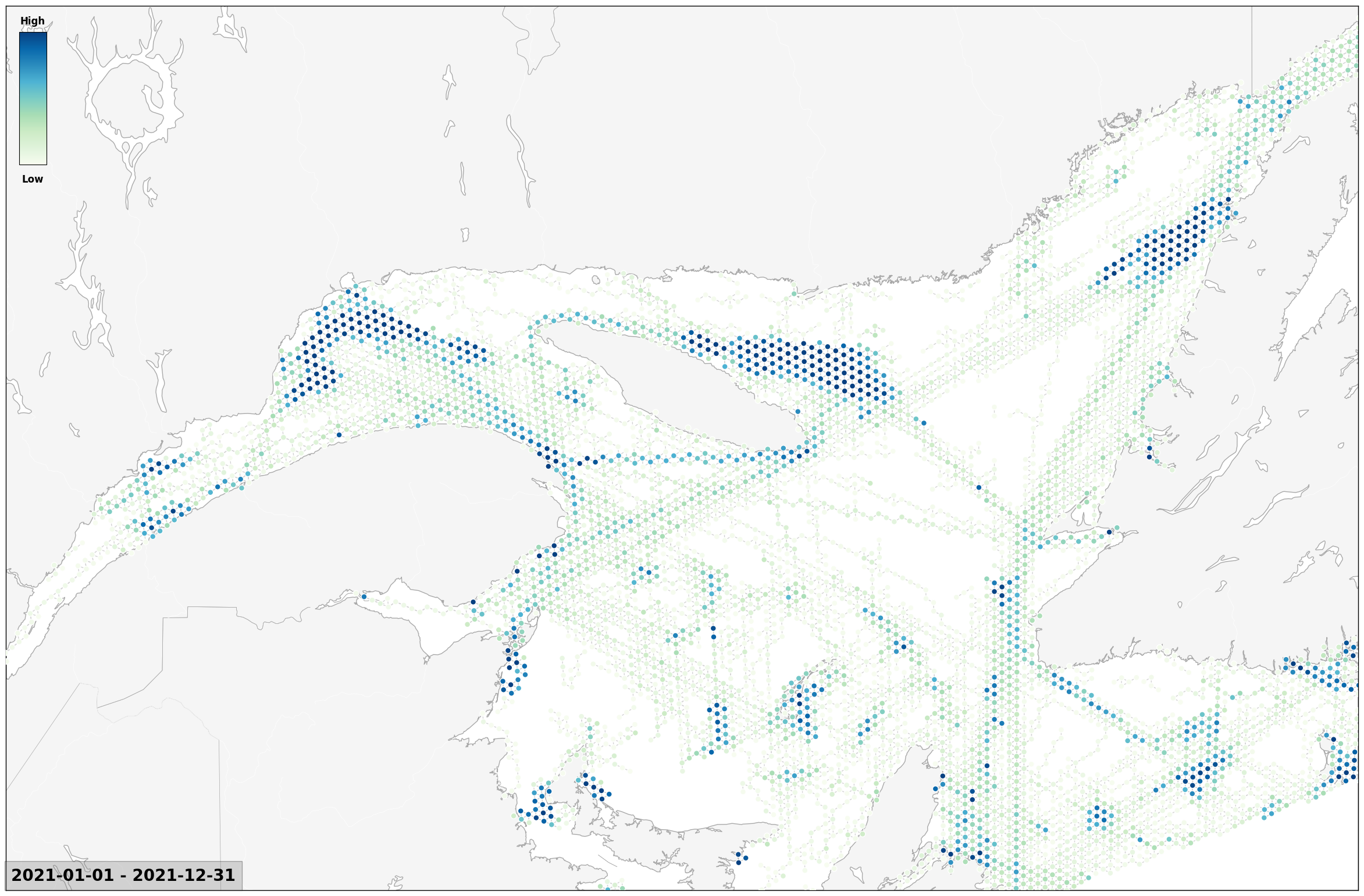} &
        \includegraphics[width=0.33\linewidth]{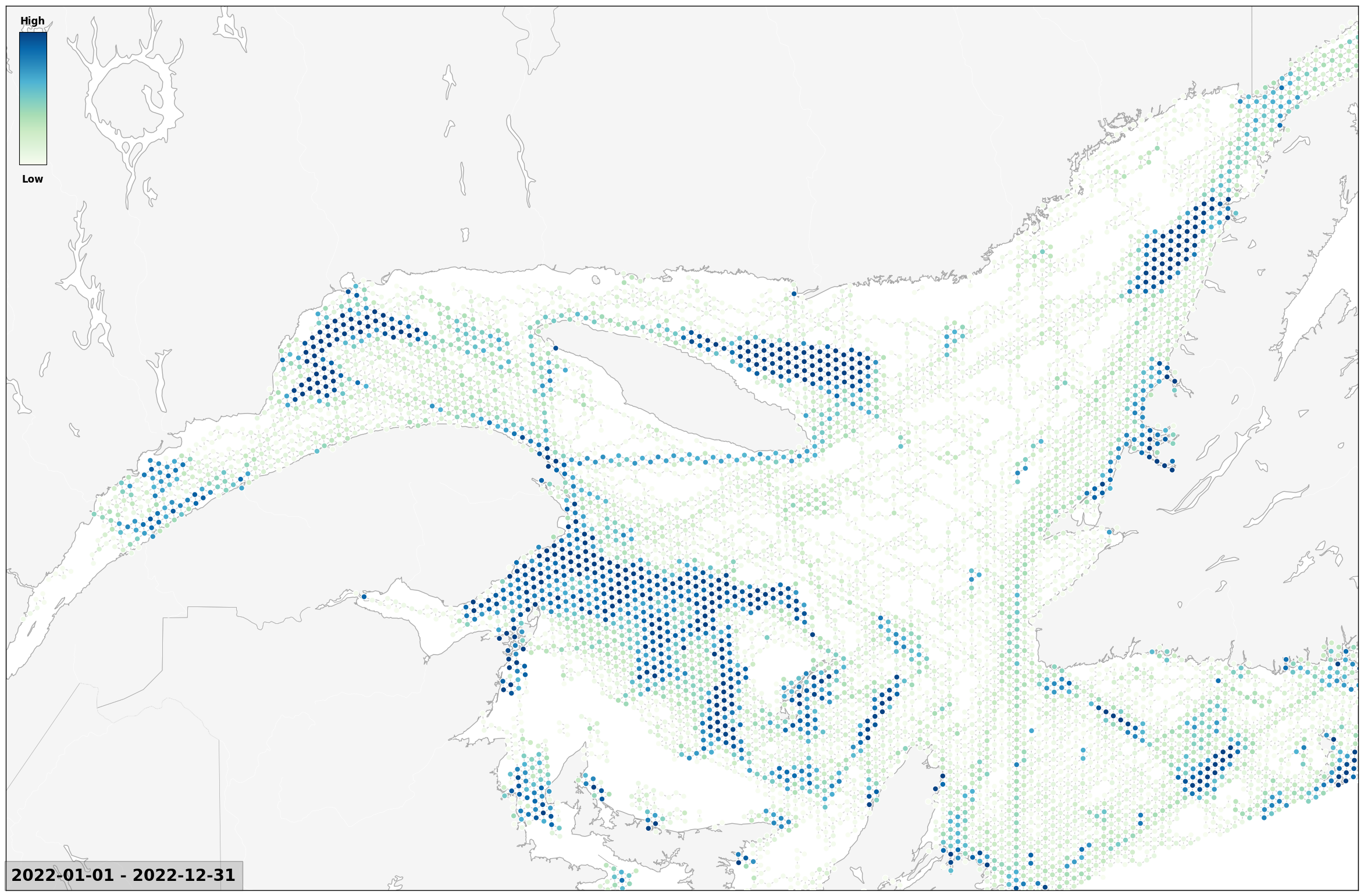} \\
        \raisebox{1\height}{\rotatebox[origin=c]{90}{\hspace{1.1cm}\textit{\textbf{Passenger}}}} &
        \includegraphics[width=0.33\linewidth]{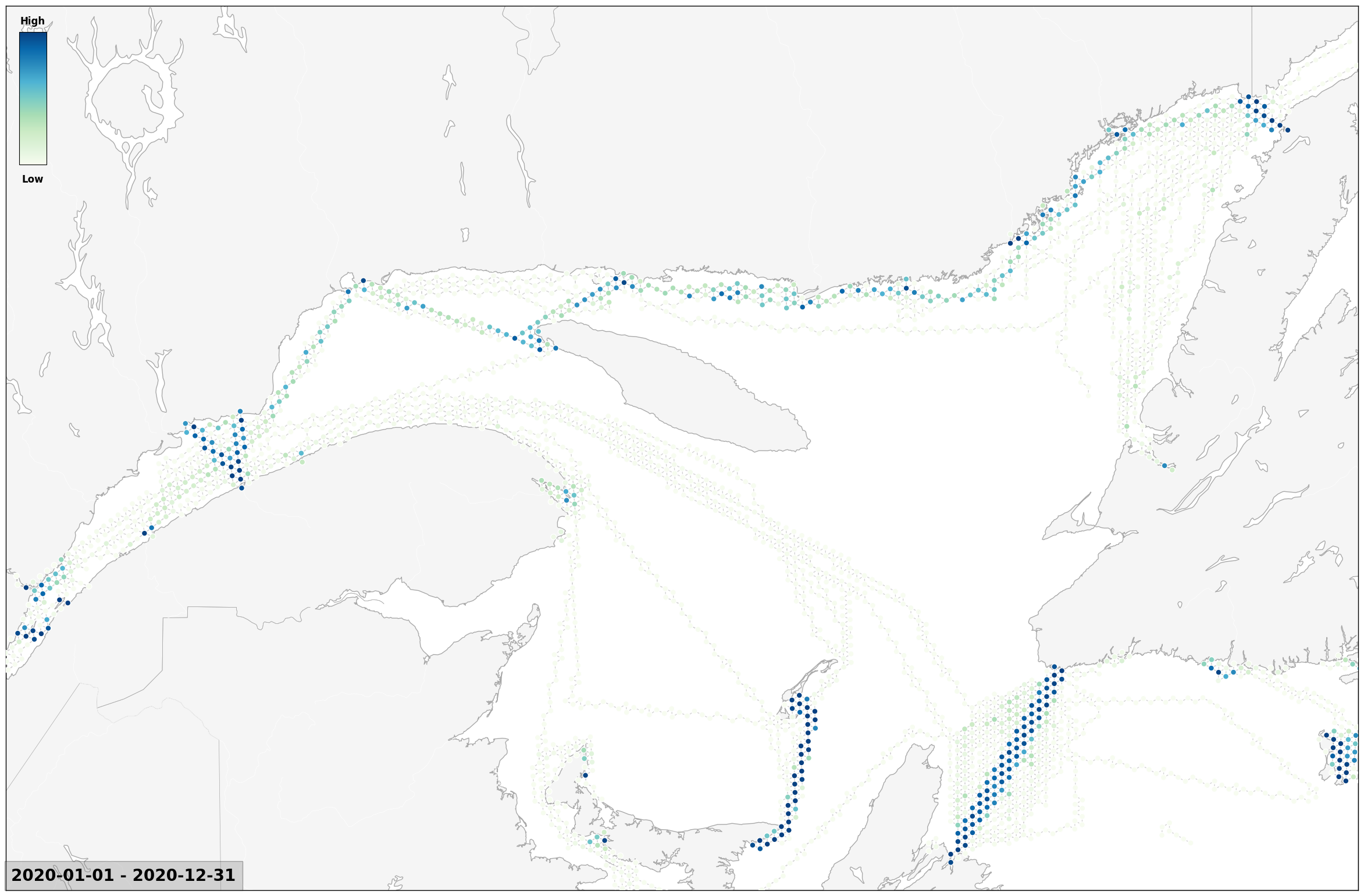} &
        \includegraphics[width=0.33\linewidth]{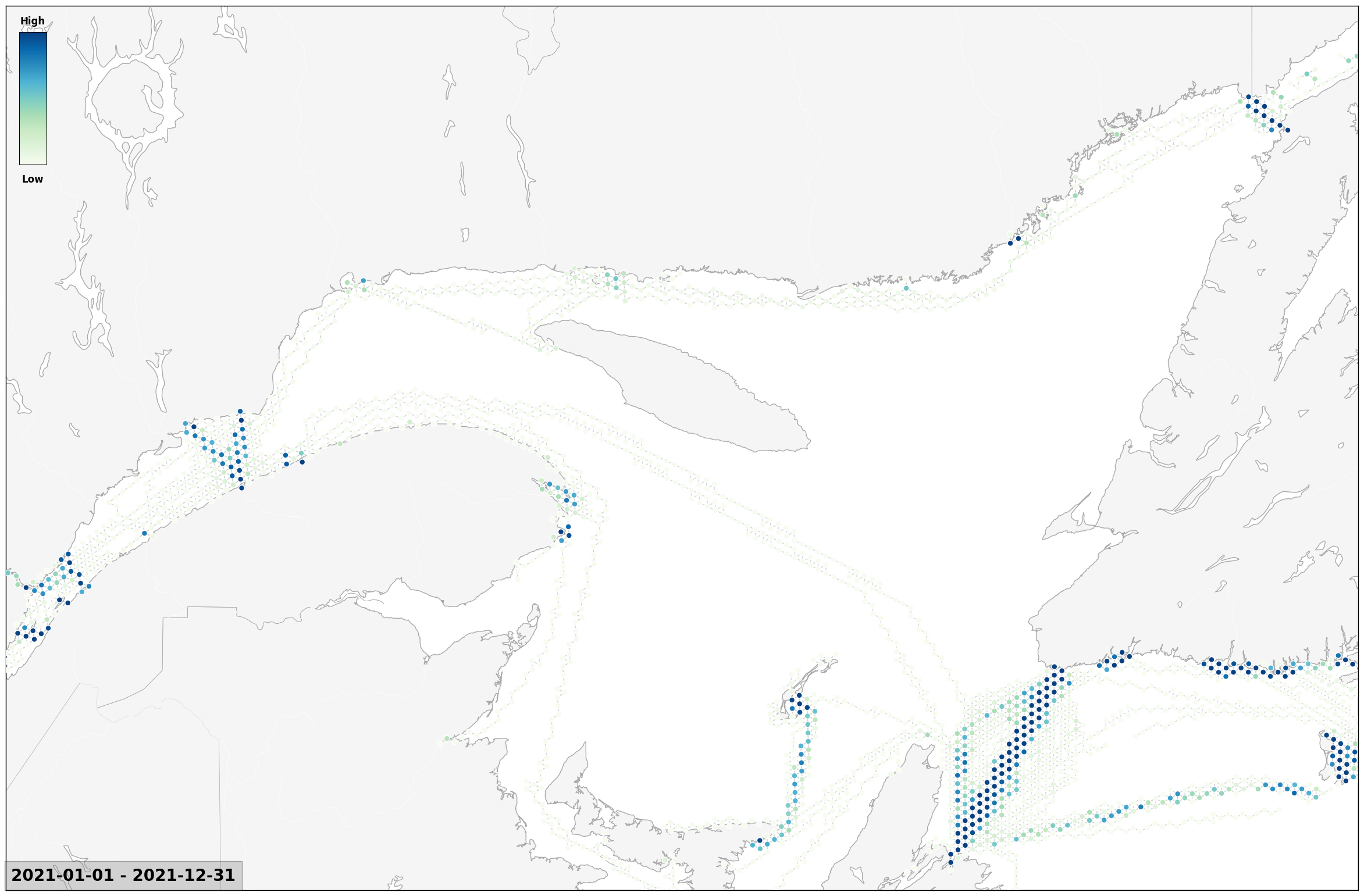} &
        \includegraphics[width=0.33\linewidth]{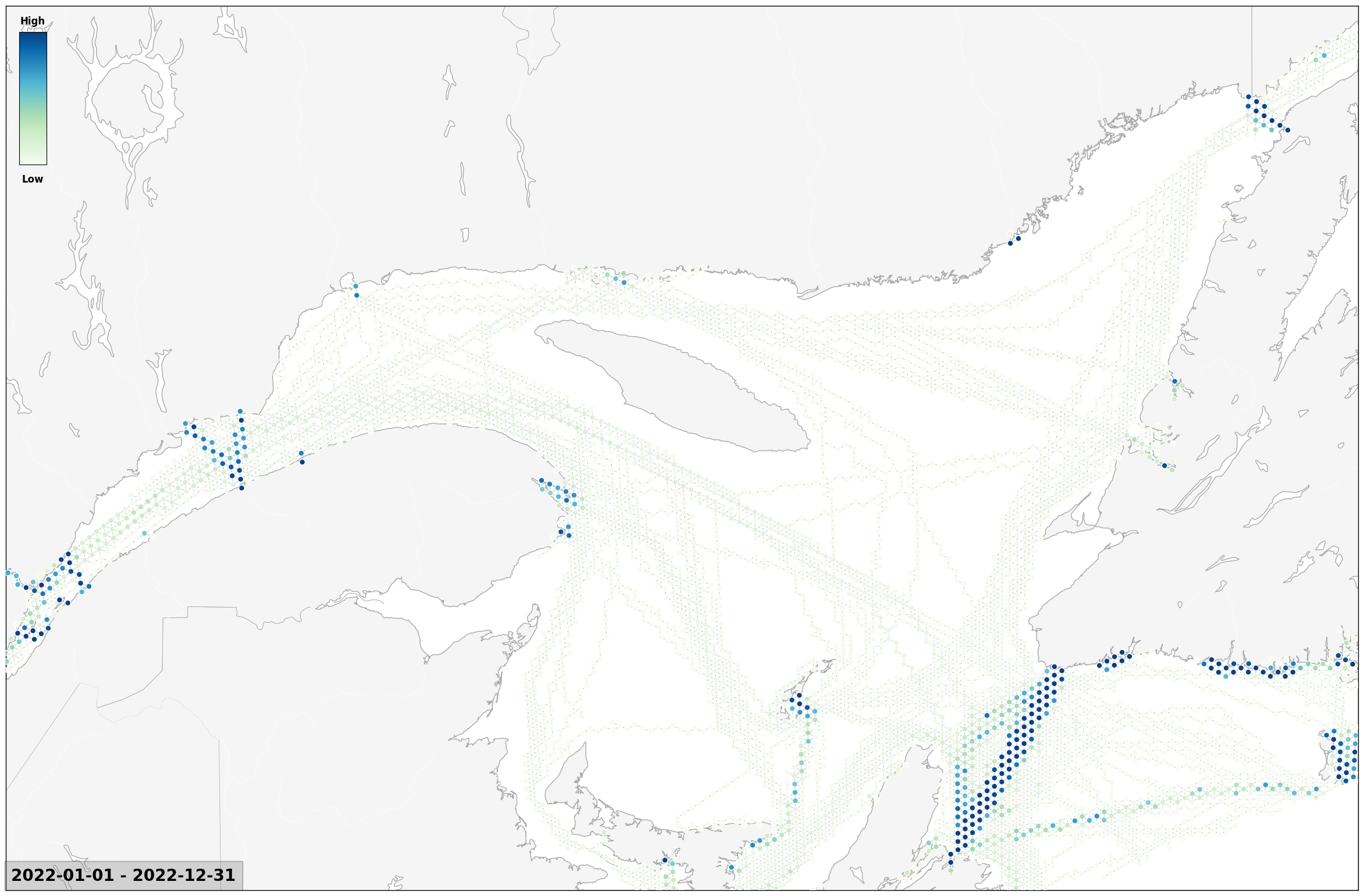} \\
    \end{tabular}
    \caption{\textbf{Raw Dwell‑Time Intensity.} Spatial footprint of accumulated dwell per cell by vessel type across 2020--2022.}
    \label{fig:dwell}
\end{figure*}

\vspace{.1cm}
\noindent\textit{\textbf{Pandemic‑Driven Maritime Traffic Redistribution}}
\vspace{.1cm}

Figures~\ref{fig:transition} and~\ref{fig:dwell} extend the preceding spatial analysis by presenting raw transition and dwell counts across vessel types and years. Building on the accumulative cell-level metrics from the pre-pandemic baseline in Figure~\ref{fig:transition_dwell_2019}, these results retain absolute values and apply the transformation in Equation~\ref{eq:clip_and_transform}, enabling the identification of high- and low-activity regions.

Overall, the spatial extent of vessel activity progressively expands over time, as evidenced by the increasing number of active grid cells. Transition coverage rises from approximately 17\% of cells in 2020 to 20\% in 2022, while dwell coverage increases from 16\% to nearly 19\%, indicating a steady recovery in maritime activity following pandemic-related restrictions. The fishing sector primarily drives this expansion, with transition coverage growing from 14\% to 19\% and dwell coverage increasing from 12\% to 15\%. These trends underline the resilience and spatial growth of the fishing industry during the post-pandemic period.

In contrast, commercial vessels exhibit only marginal changes in spatial coverage, suggesting that shipping volumes remained largely stable along established maritime corridors throughout the pandemic and subsequent recovery. Passenger vessels continue to operate within tightly defined transition zones centered around terminal routes. However, many routes show notable temporal variations due to restrictive travel policies during the pandemic, including temporary suspensions.

Fishing vessel transitions (Figure~\ref{fig:transition}) in 2021 and 2022 reveal the emergence of new high-frequency corridors not observed in the pre- or early pandemic periods. Two routes become increasingly prominent: an eastbound arc along the southern coast of Anticosti Island and a northbound diagonal extending from Cape Breton to the Lower North Shore. In contrast, dwell hotspots for fishing vessels (Figure~\ref{fig:dwell}) shift away from the coastal Gaspé region toward offshore zones east of the Laurentian Channel. This redistribution indicates a post-pandemic realignment in fishing operations, with increased activity in deeper and less congested areas.

Passenger vessel activity reflects pandemic-induced disruptions. Figure~\ref{fig:transition} shows a complete absence of passenger transitions along the Argentia $\Leftrightarrow$ North Sydney route in 2020, corresponding to the full suspension of Marine Atlantic’s seasonal ferry service under COVID-19 regulations~\cite{marineatlantic2020}. A similar gap is observed along the Les Escoumins–Trois-Pistoles route, which was canceled for the entire 2020 season in response to public health measures~\cite{ferryservice2019}.

\section{Conclusions}
\label{sec:conclusions}

This paper presented a Markov-chain framework for modeling maritime mobility using AIS trajectories, enabling consistent, scalable, and interpretable analysis of vessel behavior across time and vessel types. By discretizing the maritime domain into spatial grids and modeling transitions and dwell-time distributions within a stochastic process formulation, we derived mobility metrics that offer both spatial resolution and systemic insight. Applied to the Gulf of St. Lawrence, the proposed framework revealed long-term structural maturation of the maritime network, vessel-type-specific operational behavior, and temporal disruptions induced by the pandemic.

The stabilization of state-space size and connectivity after 2018, accompanied by increasing modularity and declining average path lengths, indicates that the AIS-derived mobility network reached a representative level of operational completeness. These patterns reflect the consolidation of persistent traffic corridors and a more balanced distribution of vessel flows. As Betweenness centrality decreased and plateaued, the system transitioned away from chokepoint-dominated routing. This shift is more reflective of improved data availability and reporting compliance, rather than an actual decrease in physical congestion or navigational constraints.

The pandemic period introduced asymmetric disruptions across vessel types. Commercial vessels demonstrated a relatively rapid spatial recovery in transition activity by 2022, yet sustained dwell-time accumulation at anchorage zones suggests lingering inefficiencies in the broader logistics and port infrastructure. Fishing vessels, in contrast, exhibited simultaneous increases in both transitions and dwell-time magnitude, accompanied by a 35\% rise in spatial coverage relative to the 2019 pre-pandemic baseline. This expansion implies a sector-wide adaptation in operational footprint and resource targeting. Passenger vessels remained spatially constrained, and activity levels did not return to pre-pandemic levels, indicating longer-term impacts that may require further investigation.

Beyond retrospective mobility characterization, the framework offers practical utility for maritime governance. Transition probability matrices can inform adaptive routing to mitigate collision risk and protect sensitive ecosystems. At the same time, dwell-time intensity maps may support the identification of zones requiring environmental regulation or port management interventions. Since the metrics are computed from resampled trajectories and do not depend on manual parameter tuning, the system is well-suited for integration into real-time or near-real-time monitoring platforms.

Future work will explore the incorporation of dynamic environmental covariates, such as sea state and meteorological conditions, into the state-space formulation to better capture exogenous influences on navigational behavior. Additional extensions will consider variable-order Markov models for vessel classes exhibiting memory-dependent routing and evaluate their coupling with simulation-based forecasting methods to support operational decision-making and scenario planning.

\section*{Acknowledgment}

This research was partially supported by the \textit{Natural Sciences and Engineering Research Council of Canada} (NSERC), the \textit{Department of Fisheries and Oceans Canada} (DFO), and \textit{Dalhousie University} (DAL). The data used in this study were provided by the AISviz: Making Vessels Tracking Data and Maps Available to Everyone project from MERIDIAN (Marine Environmental Research Infrastructure for Data Integration and Application Network) and are subject to licensing restrictions, preventing the sharing of raw data.

\balance
\bibliographystyle{IEEEtran}
\bibliography{references}

\begin{thebibliography}{10}
\providecommand{\url}[1]{#1}
\csname url@samestyle\endcsname
\providecommand{\newblock}{\relax}
\providecommand{\bibinfo}[2]{#2}
\providecommand{\BIBentrySTDinterwordspacing}{\spaceskip=0pt\relax}
\providecommand{\BIBentryALTinterwordstretchfactor}{4}
\providecommand{\BIBentryALTinterwordspacing}{\spaceskip=\fontdimen2\font plus
\BIBentryALTinterwordstretchfactor\fontdimen3\font minus \fontdimen4\font\relax}
\providecommand{\BIBforeignlanguage}[2]{{%
\expandafter\ifx\csname l@#1\endcsname\relax
\typeout{** WARNING: IEEEtran.bst: No hyphenation pattern has been}%
\typeout{** loaded for the language `#1'. Using the pattern for}%
\typeout{** the default language instead.}%
\else
\language=\csname l@#1\endcsname
\fi
#2}}
\providecommand{\BIBdecl}{\relax}
\BIBdecl

\bibitem{unctad2023rmt}
{United Nations Conference on Trade and Development}, \emph{Review of Maritime Transport 2023: Towards a Green and Just Transition}, ser. Review of Maritime Transport.\hskip 1em plus 0.5em minus 0.4em\relax Geneva: United Nations, 2023.

\bibitem{paolo2024satellite}
F.~S. Paolo, D.~Kroodsma, J.~Raynor, T.~Hochberg, P.~Davis, J.~Cleary, L.~Marsaglia, S.~Orofino, C.~Thomas, and P.~Halpin, ``Satellite mapping reveals extensive industrial activity at sea,'' \emph{Nature}, no. 7993, 2024.

\bibitem{sharif2024multi}
N.~Sharif, M.~Rönnqvist, J.-F. Cordeau, J.-F. Audy, G.~Warya, and T.~Ngo, ``Multi-objective vessel routing problems with safety considerations: A review,'' \emph{Maritime Transport Research}, vol.~7, 2024.

\bibitem{harvey1999preliminary}
M.~Harvey, M.~Gilbert, D.~Gauthier, and D.~M. Reid, \emph{A preliminary assessment of risks for the ballast water-mediated introduction of nonindigenous marine organisms in the Estuary and Gulf of St. Lawrence}.\hskip 1em plus 0.5em minus 0.4em\relax Fisheries \& Oceans Canada, Regional Oceans Branch, Maurice Lamontagne Institute, 1999.

\bibitem{Stach2023Maritime}
T.~Stach, Y.~Kinkel, M.~Constapel, and H.-C. Burmeister, ``Maritime anomaly detection for vessel traffic services: A survey,'' \emph{Journal of Marine Science and Engineering}, vol.~11, no.~6, p. 1174, 2023.

\bibitem{fu2020ais}
X.~Fu, Z.~Xiao, H.~Xu, V.~Jayaraman, N.~B. Othman, C.~P. Chua, and M.~Lind, ``Ais data analytics for intelligent maritime surveillance systems,'' in \emph{Maritime Informatics}, M.~Lind, M.~Michaelides, R.~Ward, and R.~T.~Watson, Eds.\hskip 1em plus 0.5em minus 0.4em\relax Springer International Publishing, 2021.

\bibitem{Dalaklis2023}
D.~Dalaklis, N.~Nikitakos, D.~Papachristos, and A.~Dalaklis, \emph{Opportunities and Challenges in Relation to Big Data Analytics for the Shipping and Port Industries}.\hskip 1em plus 0.5em minus 0.4em\relax Springer International Publishing, 2023.

\bibitem{imo2000ais}
\BIBentryALTinterwordspacing
{International Maritime Organization}, ``Adoption of the automatic identification system (ais) under solas chapter v,'' International Maritime Organization (IMO), 2000, aIS requirement adopted in 2000, effective for all ships by December 31, 2004. [Online]. Available: \url{https://www.imo.org/en/OurWork/safety/navigation/ais.aspx}
\BIBentrySTDinterwordspacing

\bibitem{canada2019navigation}
\BIBentryALTinterwordspacing
{Government of Canada}, ``Regulations amending the navigation safety regulations, sor/2019-100,'' Canada Gazette, Part II, 2019, canada Gazette, Part II, Volume 153, Number 9. [Online]. Available: \url{https://gazette.gc.ca/rp-pr/p2/2019/2019-05-01/html/sor-dors100-eng.html}
\BIBentrySTDinterwordspacing

\bibitem{fournier2018past}
M.~Fournier, R.~C. Hilliard, S.~Rezaee, and R.~Pelot, ``Past, present, and future of the satellite-based automatic identification system: areas of applications (2004--2016),'' \emph{WMU Journal of Maritime Affairs}, 2018.

\bibitem{spadon2024probabilistic}
G.~Spadon, J.~Kumar, D.~Eden, J.~{van Berkel}, T.~Foster, A.~Soares, R.~Fablet, S.~Matwin, and R.~Pelot, ``Multi-path long-term vessel trajectories forecasting with probabilistic feature fusion for problem shifting,'' \emph{Ocean Engineering}, vol. 312, pp. 119--138, 2024.

\bibitem{hermannsen2019recreational}
L.~Hermannsen, L.~Mikkelsen, J.~Tougaard, K.~Beedholm, M.~Johnson, and P.~T. Madsen, ``Recreational vessels without automatic identification system (ais) dominate anthropogenic noise contributions to a shallow water soundscape,'' \emph{Scientific Reports}, vol.~9, no.~1, p. 15477, 2019.

\bibitem{lu2013approaching}
X.~Lu, E.~Wetter, N.~Bharti, A.~J. Tatem, and L.~Bengtsson, ``Approaching the limit of predictability in human mobility,'' \emph{Scientific Reports}, 2013.

\bibitem{song2010modelling}
C.~Song, T.~Koren, P.~Wang, and A.-L. Barabási, ``Modelling the scaling properties of human mobility,'' \emph{Nature Physics}, 2010.

\bibitem{liu2020universal}
E.-J. Liu and X.-Y. Yan, ``A universal opportunity model for human mobility,'' \emph{Scientific Reports}, vol.~10, no.~1, p. 4657, 2020.

\bibitem{Saputra2024Mobility}
R.~Saputra, Suprapto, and A.~Sihabuddin, ``Mobility prediction using markov models: A survey,'' in \emph{2024 7th Int. Conf. on Info. and Comp. Sci. (ICICoS)}, 2024, pp. 508--513.

\bibitem{farnoosh2020deepmarkov}
A.~Farnoosh, B.~Rezaei, E.~Z. Sennesh, Z.~Khan, J.~Dy, A.~Satpute, J.~B. Hutchinson, J.-W. van~de Meent, and S.~Ostadabbas, ``Deep markov spatio-temporal factorization,'' \emph{arXiv}, 2020.

\bibitem{alam2025physics}
M.~M. Alam, A.~Soares, J.~F. Rodrigues-Jr, and G.~Spadon, ``Physics-informed neural networks for vessel trajectory prediction: Learning time-discretized kinematic dynamics via finite differences,'' \emph{arXiv}, 2025.

\bibitem{wenzhe2025stgdpm}
J.~Wenzhe, T.~Haina, and Z.~Xudong, ``Stgdpm: Vessel trajectory prediction with spatio-temporal graph diffusion probabilistic model,'' \emph{arXiv}, 2025.

\bibitem{nguyen2024}
D.~Nguyen and R.~Fablet, ``A transformer network with sparse augmented data representation and cross entropy loss for ais-based vessel trajectory prediction,'' \emph{IEEE Access}, vol.~12, p. 21596–21609, 2024.

\bibitem{yu2025multimodal}
H.~Yu, T.~Li, K.~Torp, and C.~S. Jensen, ``A multi-modal knowledge-enhanced framework for vessel trajectory prediction,'' \emph{arXiv}, 2025.

\bibitem{Yan2021Mobility}
M.~Yan, S.~Li, C.~A. Chan, Y.~Shen, and Y.~Yu, ``Mobility prediction using a weighted markov model based on mobile user classification,'' \emph{Sensors}, vol.~21, no.~5, 2021.

\bibitem{Shi2024Combining}
Y.~Shi, H.~Tao, and L.~Zhuo, ``Combining the spatiotemporal mobility patterns and mmc for next location prediction of fake base stations,'' \emph{Computational Urban Science}, vol.~4, no.~1, p.~20, 2024.

\bibitem{Xia2023Community}
C.~Xia, Y.~Hu, and J.~Chen, ``Community time-activity trajectory modeling based on markov chain simulation and dirichlet regression,'' \emph{Computers, Environment and Urban Systems}, vol. 100, p. 101933, 2023.

\bibitem{Kim2022Maritime}
Y.-J. Kim, J.-S. Lee, A.~Pititto, L.~Falco, M.-S. Lee, K.-K. Yoon, and I.-S. Cho, ``Maritime traffic evaluation using spatial-temporal density analysis based on big ais data,'' \emph{Applied Sciences}, vol.~12, no.~21, 2022.

\bibitem{March2021Tracking}
D.~March, K.~Metcalfe, J.~Tintoré, and B.~J. Godley, ``Tracking the global reduction of marine traffic during the covid-19 pandemic,'' \emph{Nature Communications}, vol.~12, no.~1, p. 2415, 2021.

\bibitem{Loveridge2024Context}
A.~Loveridge, C.~D. Elvidge, D.~A. Kroodsma, T.~D. White, K.~Evans, A.~Kato, Y.~Ropert-Coudert, J.~Sommerfeld, A.~Takahashi, R.~Patchett, B.~Robira, C.~Rutz, and D.~W. Sims, ``Context-dependent changes in maritime traffic activity during the first year of the covid-19 pandemic,'' \emph{Global Environmental Change}, vol.~84, p. 102773, 2024.

\bibitem{Wang2022Quantitative}
X.~Wang, Z.~Liu, R.~Yan, H.~Wang, and M.~Zhang, ``Quantitative analysis of the impact of covid-19 on ship visiting behaviors to ports: A framework and a case study,'' \emph{Ocean \& Coastal Management}, 2022.

\bibitem{Zhang2016}
S.-k. Zhang, Z.-j. Liu, Y.~Cai, Z.-l. Wu, and G.-y. Shi, ``Ais trajectories simplification and threshold determination,'' \emph{Journal of Navigation}, vol.~69, no.~4, p. 729–744, 2016.

\bibitem{sahr2011hexagonal}
K.~Sahr, ``Hexagonal discrete global grid systems for geospatial computing,'' \emph{Archiwum Fotogrametrii, Kartografii i Teledetekcji}, 2011.

\bibitem{Wozniak2021hex2vec}
S.~Wo\'{z}niak and P.~Szyma\'{n}ski, ``hex2vec: Context-aware embedding h3 hexagons with openstreetmap tags,'' in \emph{Proceedings of the 4th ACM SIGSPATIAL International Workshop on AI for Geographic Knowledge Discovery}, ser. GEOAI '21.\hskip 1em plus 0.5em minus 0.4em\relax New York, NY, USA: ACM, 2021.

\bibitem{Guo2018kChain}
S.~Guo, C.~Liu, Z.~Guo, Y.~Feng, F.~Hong, and H.~Huang, ``Trajectory prediction for ocean vessels base on k-order multivariate markov chain,'' in \emph{Wireless Algorithms, Systems, and Applications}, S.~Chellappan, W.~Cheng, and W.~Li, Eds.\hskip 1em plus 0.5em minus 0.4em\relax Springer Int. Publishing, 2018.

\bibitem{Kulkarni2019}
V.~Kulkarni, A.~Mahalunkar, B.~Garbinato, and J.~D. Kelleher, ``On the inability of markov models to capture criticality in human mobility,'' in \emph{Artificial Neural Networks and Machine Learning -- ICANN 2019: Image Processing}, I.~V. Tetko, V.~K{\r{u}}rkov{\'a}, P.~Karpov, and F.~Theis, Eds.\hskip 1em plus 0.5em minus 0.4em\relax Cham: Springer International Publishing, 2019, pp. 484--497.

\bibitem{Asanjarani2021}
A.~Asanjarani, B.~Liquet, and Y.~Nazarathy, ``Estimation of semi-markov multi-state models: a comparison of the sojourn times and transition intensities approaches,'' \emph{Int. J. of Biostatistics}, vol.~18, 2021.

\bibitem{smith2021big}
D.~Smith, ``Big data insights into container vessel dwell times,'' \emph{Transportation Research Record}, vol. 2675, no.~10, pp. 1222--1235, 2021.

\bibitem{Liu2021}
H.~Liu, X.~Chen, Y.~Wang, B.~Zhang, Y.~Chen, Y.~Zhao, and F.~Zhou, ``Visualization and visual analysis of vessel trajectory data: A survey,'' \emph{Visual Informatics}, vol.~5, no.~4, pp. 1--10, 2021.

\bibitem{hu2014detecting}
Y.~Hu, H.~J. Miller, and X.~Li, ``Detecting and analyzing mobility hotspots using surface networks,'' \emph{Transactions in GIS}, 2014.

\bibitem{Kuntz2021}
J.~Kuntz, P.~Thomas, G.-B. Stan, and M.~Barahona, ``Stationary distributions of continuous-time markov chains: A review of theory and truncation-based approximations,'' \emph{SIAM Review}, vol.~63, no.~1, 2021.

\bibitem{Doglioli2017BetweennessMarine}
A.~Costa, A.~A. Petrenko, K.~Guizien, and A.~M. Doglioli, ``On the calculation of betweenness centrality in marine connectivity studies using transfer probabilities,'' \emph{PLOS ONE}, vol.~12, no.~12, 12 2017.

\bibitem{Cullinane2016PortCentrality}
Y.~Wang and K.~Cullinane, ``Determinants of port centrality in maritime container transportation,'' \emph{Transp. Research Part E: Logistics and Transp. Review}, vol.~95, pp. 326--340, 2016.

\bibitem{PHAC2020Isolation}
\BIBentryALTinterwordspacing
P.~H.~A. of~Canada. (2020, Mar.) New order makes self-isolation mandatory for individuals entering canada. [Online]. Available: \url{http://bit.ly/4k05JCl}
\BIBentrySTDinterwordspacing

\bibitem{waterflood2017}
\BIBentryALTinterwordspacing
W.~P. Media, ``High water levels impacting shipping industry on the st. lawrence seaway,'' July 2017. [Online]. Available: \url{https://www.wrvo.org/the-upstate-economy/2017-07-05/high-water-levels-impacting-shipping-industry-on-the-st-lawrence-seaway}
\BIBentrySTDinterwordspacing

\bibitem{gulffishing2021}
\BIBentryALTinterwordspacing
{Fisheries and Oceans Canada}, ``Fishing area maps,'' Gulf Region, 2021. [Online]. Available: \url{https://www.glf.dfo-mpo.gc.ca/glf/en/fishing-area-maps}
\BIBentrySTDinterwordspacing

\bibitem{marineatlantic2020}
\BIBentryALTinterwordspacing
{Marine Atlantic}, ``Marine atlantic to suspend argentia route for 2020 season,'' May 2020. [Online]. Available: \url{https://www.marineatlantic.ca/media-galleries/media-releases/marine-atlantic-suspend-argentia-route-2020-season}
\BIBentrySTDinterwordspacing

\bibitem{ferryservice2019}
\BIBentryALTinterwordspacing
{CBC News}, ``Trois-pistoles—les escoumins ferry service cancelled for 2020 season,'' November 2019. [Online]. Available: \url{https://www.cbc.ca/news/canada/montreal/trois-pistoles-les-escoumins-2020-ferry-service-cancelled-1.5366096}
\BIBentrySTDinterwordspacing

\end{thebibliography}

\end{document}